\documentclass{article}

\usepackage{graphicx}
\usepackage{epsfig}

\usepackage{amsmath}
\usepackage{amssymb}
\usepackage{mathrsfs}

\def\scrI{{\mathscr I}}

\def\scri{\scrI}

\def\defscript{\mathscr}

\def\B{{\defscript B}}

\def\H{{\defscript H}}
\def\I{{\defscript I}}

\def\LL{{\defscript L}}
\def\M{{\defscript M}}

\def\R{{\defscript R}}

\def\ifempty#1{\def\tmpdata{#1}\ifx\tmpdata\empty }

\def\linebreak{\hfill\break}

\def\mpl{m_{\rm pl}}

\def\sun{\odot}
\def\Msun{M_\sun}

\def\bra<#1|{\langle #1\rvert}
\def\ket|#1>{\lvert#1 \rangle}
\def\braket<#1|#2>{\langle #1|#2 \rangle}

\def\pfrac#1#2{\left(\frac{#1}{#2}\right)}
\def\const{\text{const}}
\def\otop#1{\hbox{$#1\kern-0.1em$\llap{\hbox{\raise1.7ex\hbox{$\scriptstyle\circ$}}}} }

\def\inpare#1{\left(#1\right)}
\def\bigpare(#1){\left(#1\right)}
\def\inrbra#1{\left\{ #1 \right\}}
\def\insbra#1{\left[ #1 \right]}
\def\inang#1{\left\langle {#1} \right\rangle}
\def\EXP#1{\inang{#1}}
\def\bigbra[#1]{\left[ #1 \right]}

\def\cases#1{\left\{ \begin{array}{ll}#1\end{array}\right.}

\def\tend{\rightarrow}
\def\then{\Rightarrow\quad}
\def\equivalent{\quad\Leftrightarrow\quad}
\def\therefore{\mbox{\setbox0=\hbox{X}\hbox{$\ldotp$}\raise0.7\ht0\hbox{$\ldotp$}\hbox{$\ldotp$}} \quad }
\def\because{\mbox{\setbox0=\hbox{X}\raise0.7\ht0\hbox{$\ldotp$}\hbox{$\ldotp$}\raise0.7\ht0\hbox{$\ldotp$}}\kern0pt }

\def\ZR{{{\mathbb Z}}}

\def\RF{{{\mathbb R}}}

\def\upin{\hbox{\setbox0=\hbox{$\cup$} \vrule width 0.05 \wd0 height \ht0 depth 0pt \kern - 0.5\wd0 \box0 }}
\def\Frac(#1/#2){\left(\frac{#1}{#2}\right)}
\def\lsim{\stackrel{<}{\sim}}
\def\gsim{\stackrel{>}{\sim}}

\def\Tr{{\rm Tr}}
\def\tr{{\rm tr}}

\def\sdprod{\mathrel{{\setbox0=\hbox{$\displaystyle\times$}\lower0.3\wd0\hbox{$\stackrel{\box0}{\scriptstyle\sim}$}}}}

\def\w{\wedge}

\def\tosigma#1,{%
    \ifx\tmpindex\relax \def\tmpindex{#1} \let\next=\tosigma
    \else \ifnum\tmpindex=0 1 \else \sigma_\tmpindex \fi
          \ifx#1\relax  \let\next=\relax
          \else \otimes \let\next=\tosigma \def\tmpindex{#1} \fi
    \fi \next}
\def\tspb(#1){\let\tmpindex=\relax\tosigma#1,\relax,}
\def\dddot#1{\stackrel{...}{#1}{}\!\!}

\def\pd{\partial}

\def\HyperG(#1,#2;#3;#4){F\inpare{\textstyle #1,#2;#3;#4}}

\def\dual{{}*\! }

\def\Eq#1{\begin{equation} #1 \end{equation}}

\def\Eqr#1{\begin{eqnarray} #1 \end{eqnarray}}
\def\Eqrn#1{\begin{eqnarray*} #1 \end{eqnarray*}}
\def\Eqrsub#1{\begin{subequations}\Eqr{#1}\end{subequations}}
\def\Eqrsubl#1#2{\begin{subequations}
  \expandafter\ifx\csname Rlabel\endcsname \relax \label{#1}
  \else \Rlabel{#1} \fi \Eqr{#2}\end{subequations}}
\def\Bitm{\begin{itemize}}
\def\Eitm{\end{itemize}}
\def\Blist#1#2{\begin{list}{#1}{\parsep=0pt \itemsep=0pt%
  \listparindent=0pt #2}}
\def\Elist{\end{list}}
\long\def\ignore#1#2{\def\ignoreflag{#1}\long\def\tmptext{#2}
  \ifnum\ignoreflag>1 #2 \fi}

\def\FigDir{.}

\newif\ifIJMP
\IJMPfalse

\ifIJMP

\else
\def\title#1{\begin{center}\Large\bf #1 \end{center}\bigskip}
\def\author#1{\begin{center} #1\end{center}}
\def\address#1{\begin{center} #1 \end{center}}
\let\refcite=\cite
\fi

\begin{document}

\title{Axiverse and Black Hole%
\footnote{This article is basen on the lecture given at the 2011 Shanghai Asia-Pacific School and Workshop on Gravitation, and  will be published in its proceedings as a volume of the International Journal of Modern  Physics Conference series. }}

\author{Hideo KODAMA
}

\address{Theory Center, Institute of Particle and Nuclear Studies, KEK, \\
      \& \\
Department of Particles and Nuclear Physics, The Graduate University for Advanced Studies,
        1-1 Oho, Tsukuba, Ibaraki 305-0801, Japan\\
hideo.kodama@kek.jp
}
\author{Hirotaka YOSHINO}
\address{Theory Center, Institute of Particle and Nuclear Studies, KEK, \\
        1-1 Oho, Tsukuba, Ibaraki 305-0801, Japan \\
hyoshino@post.kek.jp
}

\begin{abstract}
String theory/M-theory generally predicts that axionic fields with a broad mass spectrum extending below $10^{-10}{\rm eV}$ are produced after compactification to four dimensions. These axions/fields provoke a rich variety of cosmophysical phenomena on different scales depending on their masses and provide us new windows to probe the ultimate theory.  In this article, after overviewing this axiverse idea\cite{Arvanitaki.A&&2010}, I take up the black hole instability as the most fascinating one among such axionic phenomena and explain its physical mechanism and astrophysical predictions.
\end{abstract}

\section{Introduction}

At present, superstring theory/M-theory appears to be most close to the ultimate theory of Nature. Therefore, it is crucially important to find a clue indicating that these higher-dimensional theories are really behind our universe.

Because the ultimate theory is a kind of UV completion of our low energy effective theory, its characteristic new features such as the existence of extra-dimensions in general show up in high energy phenomena. Thus, one natural approach to probe the ultimate theory is to study phenomena at the high energy end experimentally. Collider experiments like  LHC and ILC are examples, but it is very difficult to raise the maximal energy to string scales in the near future. From this respect, it is more promising to probe inflation through CMB and gravitational waves, and look for cosmological relics from the early universe such as dark matter, gravitational wave background and cosmic strings, directly or indirectly.

We can also probe the ultimate theory through low energy experiments. For example, if the size and structure of the extra dimensions is not completely stabilized, the values of fundamental constants may vary in time or spatially on cosmological scales\cite{Damour.T&Polyakov1994,Witten.E1984a,Chiba.T2001A,Dent.T&Stern&Wetterrich2008}.  Further, there may exist light moduli fields.  If they contains a scalar field,  it mediates a new force with a range corresponding to the Compton wavelength $1/\mu$.  In fact, lots of experiments have been done to look for such a new force in submm ranges, although only upper bound have been obtained so far.\cite{Geraci.A&&2008}

In contrast, if these light moduli fields are pseudo-scalar, i.e., axions, it is difficult to detect them by new force search experiments, because the force mediated by such a field is proportional to the spin or velocity of matter source and decreases faster than $1/r^2$ with the distance $r$ from the source even when $r$ is shorter than  $1/\mu$.\cite{Dimopoulos.S&Giudice1996}  Hence, the mass of the axion can be very small. In particular, the Compton wavelength $1/\mu$ can become of the order of astrophysical objects or cosmological scales. In such a case, axion fields may provoke cosmophysical phenomena, as systematically discussed by Arvanitaki et al in the axiverse paper [\refcite{Arvanitaki.A&&2010}]. In this article, we take up the superradiant instability of black holes and astrophysical phenomena provoked by it as the most fascinating one among various new phenomena provoked by superlight axions.

  This article is organized as follows. First, in the next section, we briefly overview the axiverse idea and the cosmophysics based on it. Then, we focus on the black hole problem.  After reviewing the basics on black holes, we discuss the superradiance by a rotating black hole, the superradiance instability of an axionic field and its astrophysical implications in order.

\section{String Axiverse}

As is well-known, axion was first introduced into physics as a pseudo-Goldstone boson for the Peccei-Quinn symmetry to resolve the strong CP problem\cite{Peccei.R&Quinn1977}. This QCD axion was originally assumed to have a mass of MeV order and interact rather strongly with particles in the standard model. However, it was soon recognized that the existence of such a particle contradicts experiments unless its coupling to quark unless unless its coupling to the SM sector is extremely small, i.e., it is practically "invisible".\cite{Abbott.L&Sikivie1983}

The basic features of this invisible axion are summarized as follows.\cite{Kuster.M&Raffelt&Beltran2008B}
\Bitm
\item[1.] a neutral P- and CP-odd scalar coupled very weak to matter :
\Eqrsubl{QCDaxion:interactions}{
&& g_{aq}\,  a\, (\bar q \gamma_5 q):\quad
   g_{aq} \approx m_q/f_a;\ f_a\gsim 10^9 {\rm GeV},\\
&& g_{a\gamma}\, a\,  F\w F :\quad
   g_{a\gamma}\approx 1/f_a
}
\item[2.] Small mass by the QCD instanton effect:
$
m_a \sim 10^{-3} {\rm eV}  (10^{10}{\rm GeV}/ f_a)
$
\item[3.] Dark matter candidate:  $ \Omega_a= 0.01 (f_a/10^{10}{\rm GeV})^{1.175}.$
\Eitm
Here, note that the interactions \eqref{QCDaxion:interactions} have shift symmetry, i.e.  they are invariant under the transformation $a\tend a+\const$ mod. field equations. This feature has a crucial importance in protecting the axion to get a large mass by various quantum corrections.  On the basis of this, the concept of the axion is now generalized to include all pseudo-scalar particles/fields that have P and CP violating interactions with shift symmetry at the tree level.

\subsection{String axions}

\subsubsection{Origin}

The axiverse idea is based on the fact that axions defined above are expected to  be produced abundantly  by realistic compactifications of string theory/M-theory\cite{Svrcek.P&Witten2006,Arvanitaki.A&&2010}. They come form fields contained in any string theory as essential ingredients\cite{Polchinski.J1998B}.

For example, the bosonic sector of the heterotic SST  contains, in addition to the spacetime metric $g$, a dilaton $\phi$ and gauge fields $A$,  a 2-form potential $B$ whose action in the string frame reads
\Eq{
2\kappa_{10}^2 S_B = \int_{M_{10}}  -\frac{1}{2}e^{-2\phi} \dual H\w H;\quad
H= dB - \frac{\alpha'}{4}\inpare{\omega^{\rm G}_{\rm CS} - \omega^{\rm L}_{\rm CS}},
\label{HetSST:action:B}
}
where  $\alpha'$ is the inverse of the string tension, and $ \omega^{\rm G}_{\rm CS}$ and  $ \omega^{\rm L}_{\rm CS}$ are the Chern-Simons connections for the gauge field $A$ and the gravitational connection $\omega$, respectively:
\Eq{
d \omega^{\rm G}_{\rm CS}=\Tr\inpare{F\w F},\quad
d\omega^{\rm L}_{\rm CS} = \tr \inpare{\R\w\R}.
}
This action is invariant under the gauge-transformation
\Eq{
\delta A=d\lambda,\quad
\delta \omega=d\Theta,\quad
\delta B = d\sigma + (\alpha'/4)\inrbra{\Tr(\lambda dA)+\tr(\Theta d\omega)}
}
with an arbitrary 1-form $\sigma$.

When the theory is compactified on a Calabi-Yau 3-fold $Y$ to a four-dimensional spacetime $X$, the form field $B$ produces two types of axionic fields in $X$.  To see this, let us consider a simple product-type compactification, $ds^2(M_{10})= ds^2(X_4) + ds^2(Y_6)$. Let $\eta^i$($i=1,\cdots,b_2(Y)$) be a basis of harmonic 2-forms on $Y$ dual to a basis of $H_2(Y,\ZR)$. Then, $B$ can be expanded as
\Eq{
B=\ell_s^{2} \sum_{i=1}^{b_2(Y)} \alpha_i(x) \eta^i + \beta(x) ,
}
where $\ell_s=2\pi \sqrt{\alpha'}$ and $\beta(x)$ is a 2-form on $X_4$. By inserting this to the original action $S_B$, we obtain
\Eqr{
2\kappa_{10}^2 S_B = -\frac{V_Y}{2g_s^2}\int_{X_4} &&
 \big[ \sum Y^{ij} \dual d\alpha_i \w d\alpha_j + \dual h \w h
 \notag\\
&&  + \frac{\theta}{\pi} \inrbra{dh-\ell_s^2(4\pi)^{-2}\inpare{\Tr(F\w F)- \tr(\R\w\R)}}\big],
}
where $Y^{ij}=\ell_s^4 V_Y^{-1}\int_{Y_6} \dual \eta^i \w \eta^j$, $V_Y$ is the volume of $Y$, $h$ is the four-dimensional part of the 3-form $H$ and $\theta(x)$ is a Lagrange multiplier for the anomaly cancellation condition, i.e., the Bianchi identity for $h$.

From the variation with respect to $h$, we obtain $d\theta=2\pi \dual h$. By eliminating $h$ by this relation, $\theta$ is promoted to a dynamical pseudo scalar field with the action
\Eqr{
S_a= \int_{X_4} && \Big[-\frac{1}{2}\sum Y^{ij}\dual da_i\w da_j
    -\frac{1}{2} \dual da \w da
  \notag\\
 && + \frac{\lambda}{f_a} a \inrbra{\Tr(F\w F)-\tr(\R\w\R)}\Big],
\label{HetSST:action:axion}
}
where $f_a$ is the axion decay constant defined by
\Eq{
f_a=\frac{\sqrt{V_Y}}{2\sqrt{2}\pi \kappa_{10}g_s} = \frac{L^3}{\sqrt{2\pi}g_s\ell_s^4}
 =\frac{\mpl}{2\sqrt{2}\pi},
}
with $V_Y=L^6$, $a_i$ and $a$ are dimensionful axion fields defined by $a_i= f_a \alpha_i$ and $a=f_a\theta$, and $\lambda$ is a dimensionless constant
\Eq{
\lambda=\frac{\ell_s^2 f_a^2}{2\pi^2}=\frac{\mpl^2\ell_s^2}{16\pi^3}.
}
Here, note that real axion scales for $a_i$ are in general smaller than $f_a$ because $Y^{ij}\sim (\ell_s/L)^4$ generally.

Thus, we obtain two types of scalar fields, the so-called model-dependent axions $a_i$ from the internal 2-cycles and the so-called model-independent axion $a$ from $B_{\mu\nu}$. It is clear that the action \eqref{HetSST:action:axion} has shift symmetry for both types of fields. Further, the model-independent axion $a$ has pseudo-scalar couplings to the gauge and gravitational Chern-Simons terms like the QCD axion. In contrast, the model-dependent axions have apparently no such coupling and further, they are naively CP-even scalar because it is natural to regard the $B$ field as a CP-even field in the heterotic theory. However, if we consider the quantum corrections, the Green-Schwartz counter term
\Eq{
S=\int_{M_{11}} B \w X_8(F,\R)
}
produces pseudo-scalar couplings of the model-dependent axions to $\Tr(F\w F)$ and $\tr(\R\w\R).$\cite{Svrcek.P&Witten2006}

$B$ produces a model-independent axion $a$ in type IIB theory and model-dependent axions $a_i$ in type IIA theory as well, while $a_i$ fields in IIB and $a$ field in IIA become CP even. In contrast to the heterotic case, these fields do not couple to $F\w F$ or $\R\w\R$, However, in type II theories,  we have various RR form fields $C_p$, among which $C_3$  in IIA and $C_{2q}$($q=0,1,2$) in IIB theory produce model-dependent axions in the same way as those in the heterotic theory. Further, these may be coupled to gauge fields via the Chern-Simons coupling of the RR fields with gauge fields on D-branes\cite{Svrcek.P&Witten2006}.\footnote{It appears that these axions do not have a coupling to $\R\w\R$ at least in the lowest order in $\alpha'$. This feature might be used to probe the background string theory of our universe through axions.}

The very important feature of these model-dependent axions is that they are as abundant as the non-trial internal cycles. Because the number of such cycles can be very huge in the flux compactification of the type IIB theory leading to the landscape, a huge number of different axions are expected to be produced in such models\cite{Douglas.M&Kachru2007}. Although there have been proposed no definite argument on the Betti number $b_2(Y)$ in the heterotic models, systematic searches of CY 3-folds in the toric framework indicate that $b_2(Y)$ is vary large for a generic CY.\cite{Kreuzer.M2010}

\subsubsection{Mass spectrum}

If the shift symmetry is not violated at the tree level by flux, branes and compactification (i.e., by moduli stabilization),  it can be preserved by perturbative quantum corrections (for supersymmetric states). Then, axions acquire mass only by non-perturbative effects (possibly associated with SUSY breaking), such as instanton effects as in the case of the QCD axion. If a light QCD axion really exists, it is natural that there survive lots of other light axions coming from the large number of non-trivial cycles in extra-dimensions as discussed above.

Now, let us give a rough estimate of the mass of such axions. In general, the action of an axion whose mass is generated by instant effects can be written
\Eq{
\LL = -\frac{1}{2}f_a^2 (\pd\theta)^2 - \Lambda^4 U(\theta);\quad
\Lambda^4\approx M^4 e^{-S},
}
where $S$ is the instanton action. From the relations
\Eq{
m_{\rm pl}^2 \sim g_s^{-2} L^6 l_s^{-8},\quad
f_a^2 \sim g_s^{-2}  L^6 l_s^{-4} (L^2)^{-2}=g_s^{-2}L^2 l_s^{-4},\quad
S\sim l_s^{-2}L^2
}
it follows that $f_a=\mpl/S$. Hence, we have
\Eq{
m_a \approx \Lambda^2/f_a \sim (M^2/\mpl) S e^{-S/2}
}

For the QCD axion, the total potential is the sum of the QCD contribution and the stringy contribution given above:
\Eq{
V=V_{\rm QCD}+\Lambda^4 \cos\inpare{\frac{a}{f_a}+\psi};\quad
V_{\rm QCD} = \frac{a^2}{8f_a^2}r^2 F_\pi ^2 m_\pi ^2 \frac{m_u m_d}{(m_u+m_d)^2}.
}
Requiring this stringy effect to be less than the QCD instanton effect leads to the constraint
\Eq{
a \approx \frac{M^4e^{-S}}{m_\pi^2 F_\pi^2}
< 10^{-10}
\then
S\approx 200
\then
  f_a\approx 10^{16}{\rm GeV},\quad
    m_a \lsim 10^{-15}{\rm eV}
}
Thus, it is expected that there are lots of superlight axions whose mass spectrum is homogeneous in $\log m$, producing the axiverse.

\begin{figure}
\centerline{\includegraphics[width=12cm]{\FigDir/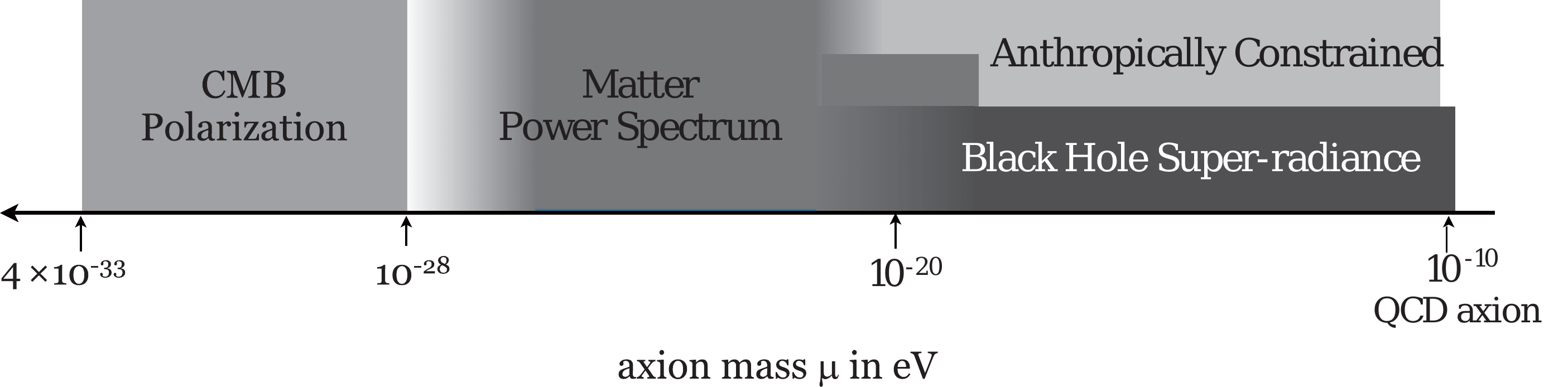}}
\vspace*{8pt}
\caption{Various axiverse windows.}
\label{fig:AxiverseSpectrum}
\end{figure}

\subsection{Axion cosmophysics}

As we saw above, typical masses of stringy axions are expected to be much smaller than the standard QCD axions. This smallness of the axion mass $\mu$ makes the Compton wavelength $1/\mu$, i.e., the minimum coherence length, comparable to cosmophysical scales and opens up possibilities for string axions to provoke the following interesting cosmophysical phenomena:\cite{Arvanitaki.A&&2010}
\Blist{$\bullet$}{}
\item Birefringence of  CMB polarization
\item Step structures in the cosmological power spectrum
\item Black hole instability/bosenova
\item Circular polarization of  primordial  GWs
\item Penetration of the GZK type barrier of  CMB for high energy gamma rays.
\Elist{}{}

For example, a coherent axionic field acts as a quintessence-type dark energy when $\mu\ll 3H$ where $H$ is the cosmic expansion rate, while it behaves as cold dark matter when $H$ becomes smaller than $\mu/3$. Typical mass scales are
\Eq{
\mu=3H= 4.5\times 10^{-33}{\rm eV} (H/H_0)
}
where $H_0$ is the present Hubble constant. The masses corresponding to $3H$ at the CMB last scattering and the radiation-matter equipartition time are around $10^{-28}{\rm eV}$ and $10^{-27}{\rm eV}$, respectively. Hence, if an axionic field with mass $\mu\gsim 10^{-28}{\rm eV}$ comprise a non-negligible fraction of dark matter, it produces a step-function-type deformation of the CDM perturbation power spectrum\cite{Arvanitaki.A&&2010,Marsh.D&Ferreira2010}, which may be observable by future experiments. When an axionic field with $5\times 10^{-33}{\rm eV}\lsim\mu\lsim 10^{-28}{\rm eV}$ is a non-negligible component, it produces rotations of the polarization of CMB\cite{Harari.D&Sikivie1992} which can be observed by the on-going and future B-mode CMB experiments.

\begin{figure}
\centerline{\includegraphics[height=7cm]{\FigDir/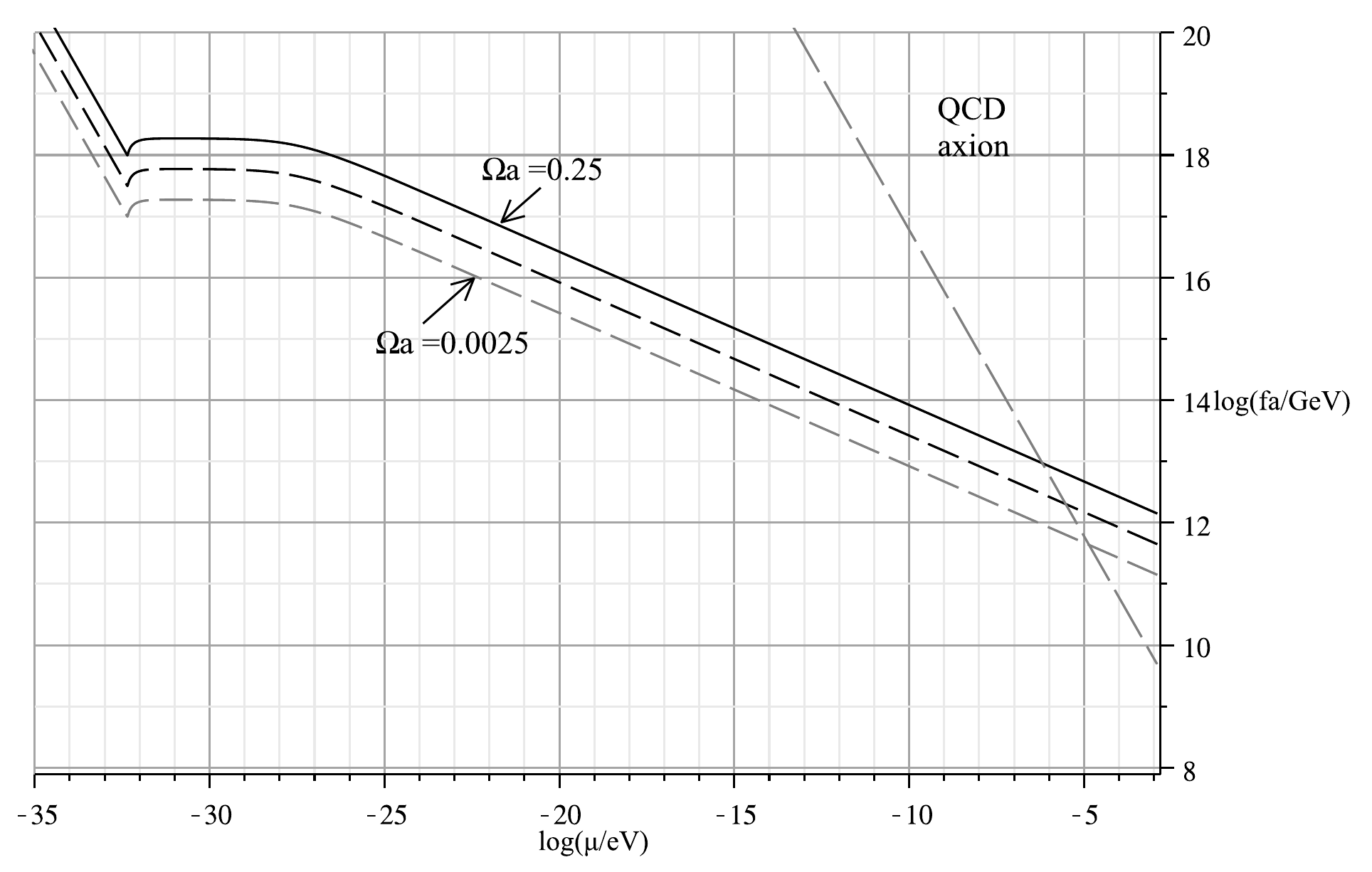}}
\caption{Density parameter of axions as a function of $\mu$ and $f_a$}
\label{fig:AxionOmega}
\end{figure}

Here, note that for string axions, there exists no universal relation between the axion mass $\mu$ and the axion decay constant $f_a$ unlike the QCD axion. Hence, we can treat them as independent parameters. Then, if we neglect corrections on $\mu$ by interactions with matter such as the temperature correction, the present density parameter $\Omega_a$ is less than the DM bound $\Omega_{\rm DM}\lsim0.25$ for a wide parameter region as shown in Fig. \ref{fig:AxionOmega}. The axion abundance in this figure is calculated under the assumption that the axion field $a$ has an amplitude of the order of $f_a$ initially, i.e., during inflation, and no dilution occurs after reheating. Hence, if we assume that the initial value of $a$ in our observed region is much smaller than that typical value for the anthropic reason or else,  the present abundance can be much smaller than the value in the figure. For example, if $f_a$ is around $10^{16}{\rm GeV}$, as suggested by the argument in the previous subsection, the mass range $10^{-20}{\rm eV} \lsim \mu \lsim 10^{-16}{\rm eV}$ is allowed only under this anthropic assumption. For $\mu \gg 10^{-16}{\rm eV}$ for which $3H$ becomes smaller than $\mu$ before BBN, dilution by decay of heavy moduli may reduce the present abundance of the axion.


Another important mass range comes from the horizon size of astrophysical black holes. As we see later, an axion field around a rotating black hole becomes unstable if its Compton wavelength is comparable to the black hole horizon size:
\Eq{
\mu \approx \frac{1}{GM} \simeq 1.3\times 10^{-10}{\rm eV}\pfrac{\Msun}{M}.
}
Because the mass $M$ of astrophysical black holes is in the range  $\Msun \lsim M \sim 10^{10}\Msun$, axions with mass in the range $10^{-20}{\rm eV}\lsim \mu \lsim 10^{-10}{\rm eV}$ can really produce such instability around black holes in binary systems and at galactic centers. In the following sections, we study this problem in more details.

\section{Black Hole Basics}

In this section, we briefly overview the basic concepts on black holes that are relevant to the superradiance instability.

\subsection{Definition of a black hole}

\begin{figure}
\centerline{
\includegraphics[height=5cm]{\FigDir/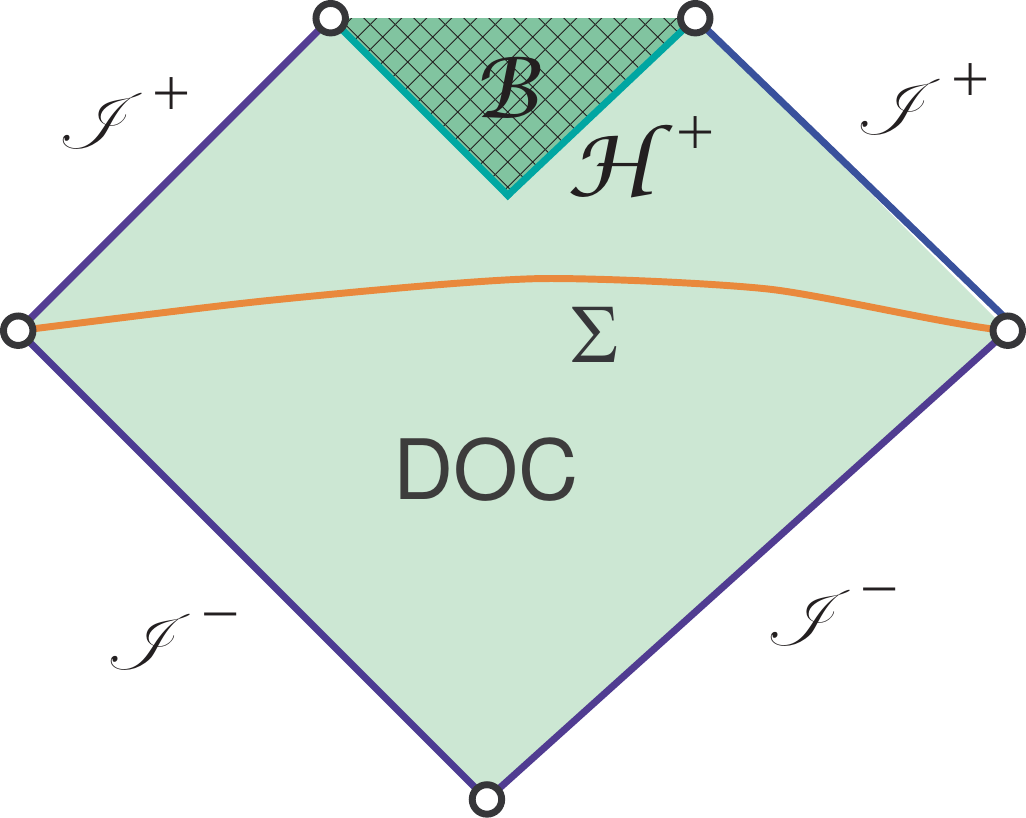}
}
\caption{Black hole spacetime}
\label{fig:BHSTgeneral}
\end{figure}

Let $\M$ be a weakly asymptotically simple spacetime and  $\scri$ be its conformal infinity\cite{Hawking.S&Ellis1973B}DIn order to avoid the appearance of singularity outside a black hole, we require that the spacetime is {\em asymptotically predictable} from a Cauchy surface $\Sigma$, i.e. $\scri \subset \overline{D(\Sigma)}\quad \text{in}\  \hat \M$. Under this condition, we define a {\em horizon} as the boundary of the region that can be observed by the infinity $\scri$ as
\Eq{
H^+= \partial(J^-(\scri))\cap J^+(\scri).
}
Then,  the {\em black hole region} is defined as the region that cannot be seen from the infinity as
\Eq{
\B= \overline{\M- J^-(\scri)},
}
and the region outside the horizon is called the {\em DOC} (Domain of outer communication):
\Eq{
{\rm DOC} = J^-(\scri,\M) \cap J^+(\scri,\M).
}
%

\subsection{Killing horizon}

\subsubsection{Stationary spacetime}

A spacetime $\M$ is said to be {\em  stationary} if there is a Killing vector $\xi$ that is timelike in some region. The metric of a stationary spacetime can be written
\Eq{
ds^2=- e^{2U(x)} (dt+A(x))^2 + g_{ij}(x)dx^i dx^j,
}
where $x=(x^i)$ is the spatial coordinates. The Killing vector $\xi$ can be written $\xi=\pd_t$ in this coordinate system, hence the corresponding 1-form is given by
\Eq{
\xi_*= -e^{2U}(dt+A(x)).
}
The rotation of the Killing vector is defined as
\Eq{
\dual(\xi_* \w d\xi_*)= -e^{3U} \dual_n dA.
}

A spacetime $\M$ is said to be {\em axisymmetric} if there is a Killing vector field $\eta$ whose orbits are all closed and are spacelike in some region. In this article, we mainly consider a stationary and axisymmetric spacetime. From the rigidity theorem for black holes, a stationary black hole is always axisymmetric if the spacetime is analytic.

\subsubsection{Killing horizon}

\begin{figure}
\begin{minipage}{5cm}
\centerline{
\includegraphics[width=4cm]{\FigDir/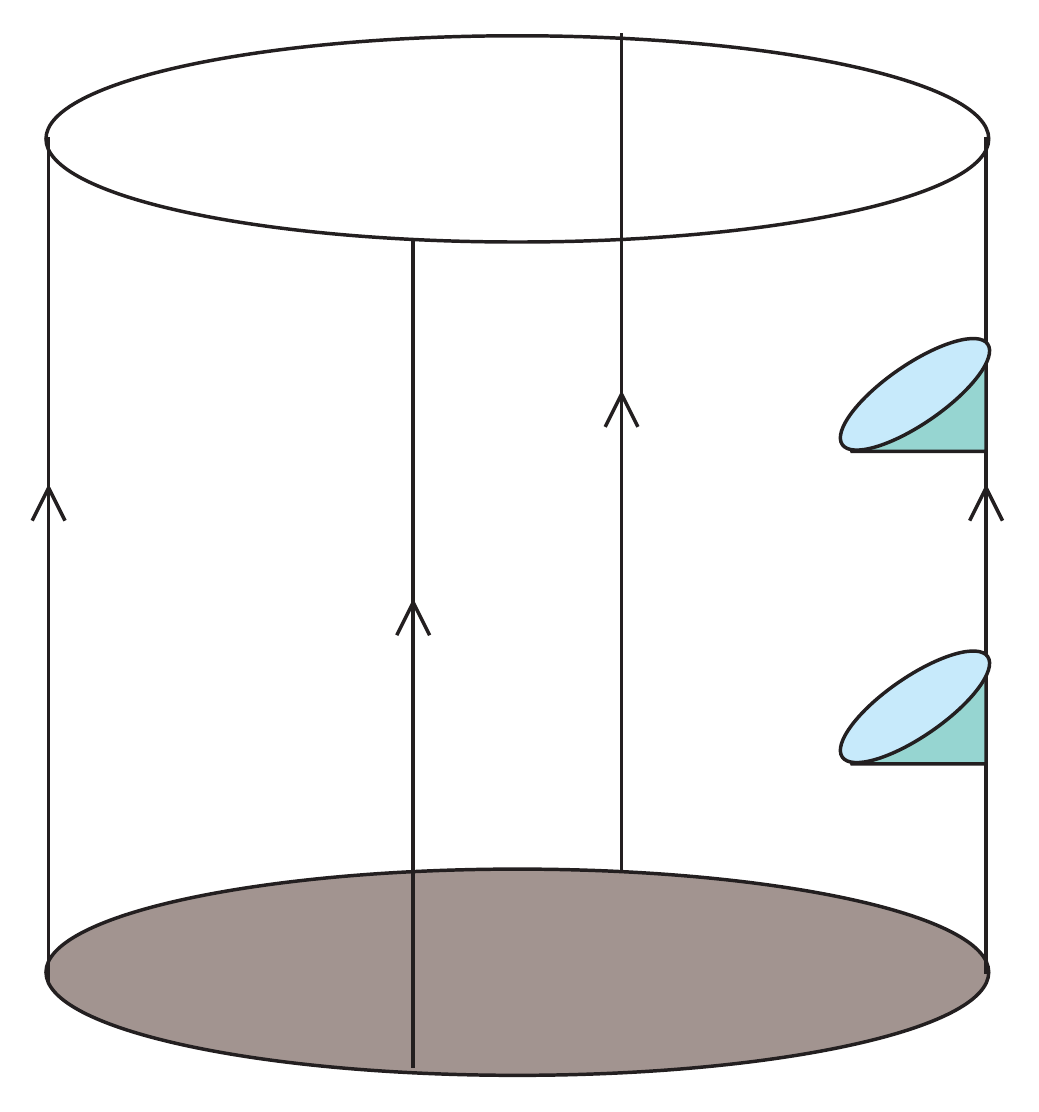}
}
\caption{Killing horizon}
\label{fig:horizongeneral}
\end{minipage}
\hspace{1cm}
\begin{minipage}{5cm}
\centerline{
\includegraphics[width=5cm]{\FigDir/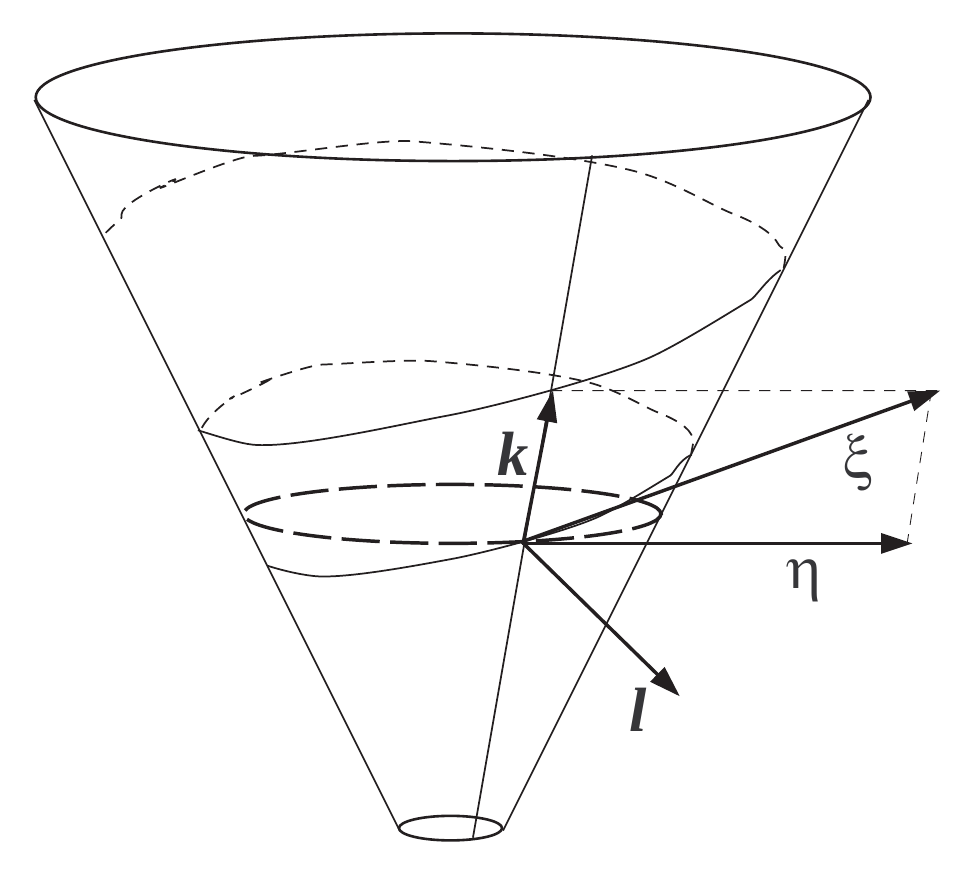}
}
\caption{Horizon of a rotating black hole}
\label{fig:rotatinghorizon}
\end{minipage}
\end{figure}

A null hypersurface $\H$ in a stationary spacetime is called a {\em Killing horizon} when there is a Killing vector that is parallel to the null geodesic generators on $\H$.
A horizon of an asymptotically simple and static spacetime with respect to infinity $\I$ is a Killing horizon if the spacetime is asymptotically predictable and the time translation Killing vector $\xi$ is timelike in a neighborhood of $\I$.

The black hole of a stationary spacetime is said to be {\em rotating} if the time-translation Killing vector is spacelike on the horizon. From the rigidity theorem, the rotating black hole horizon is a Killing horizon.

For a stationary and axisymmetric spacetime with a Killing horizon $\H$, let $\xi$ and $\eta$ be the  corresponding Killing vectors. Then, a tangent vector of the null generator of $\H$  can be uniquely written as
\Eq{
k=\xi + \Omega_h \eta.
}
$\Omega_h$ is called the {\em angular velocity of the horizon}. Further, on $\H$, we have
\Eq{
\nabla_k k = \kappa k \equivalent \nabla k^2=-2 \kappa k
}
The coefficient $\kappa$ is called the {\em surface gravity} of the black hole.

\subsection{Examples}

\begin{figure}[t]
\centerline{
\includegraphics[height=7cm]{\FigDir/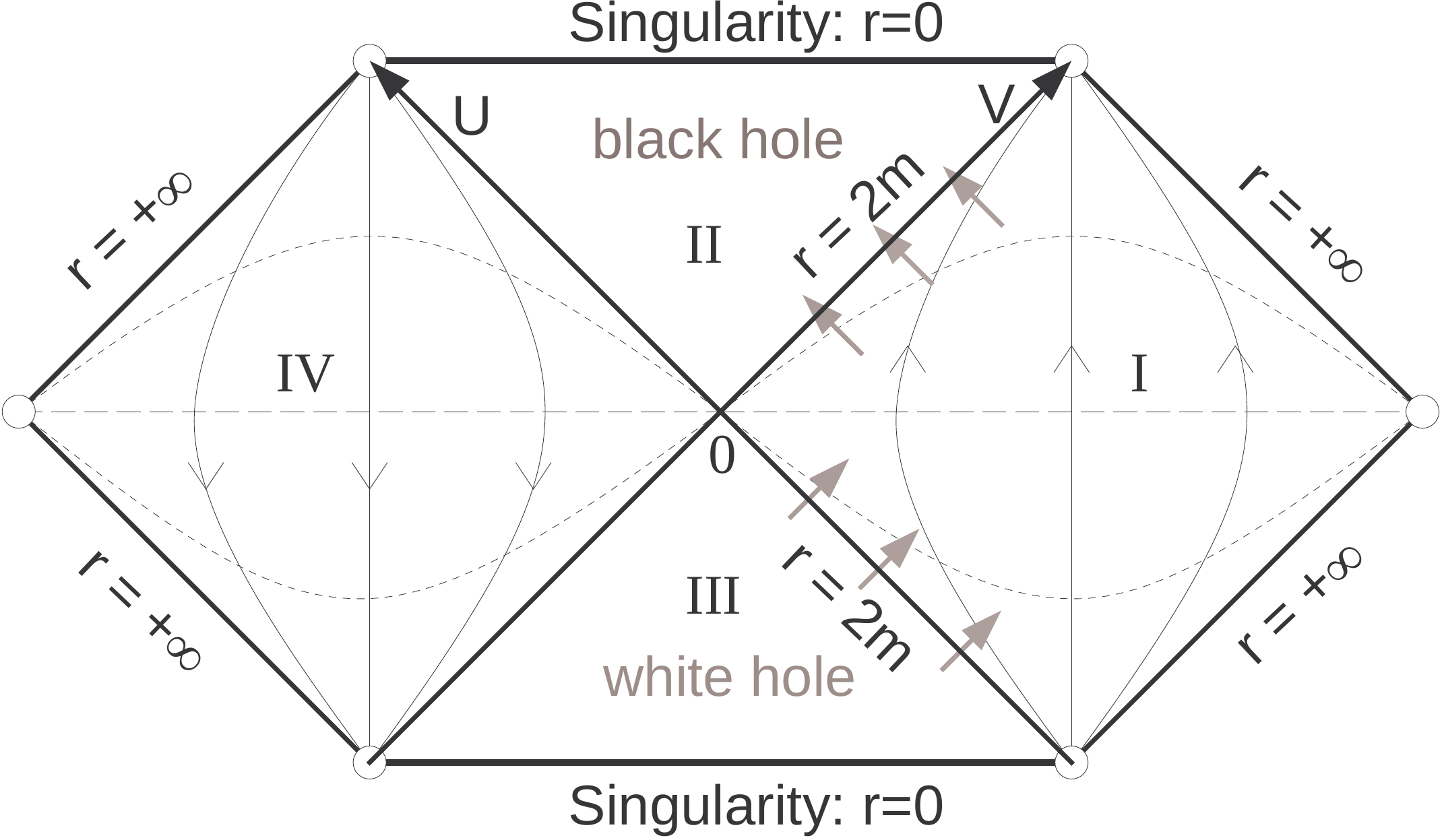}
}
\caption{Penrose diagram of the Schwarzschild black hole}
\label{fig:SSBH:penrose}
\end{figure}

\subsubsection{Static black hole}

A stationary spacetime $\M$ with the time translation Killing vector $\xi$ is said {\em static} when the rotation of $\xi$ vanishes. When a spacetime $(\M,g)$ is static, from the rotation free condition, we can find a coordinate system locally in which the metric can be written
\Eq{
ds^2=-e^{2U(x)} dt^2 + g_{ij}(x)dx^i dx^j.
}

The simplest and most important example of a static black hole is the spherically symmetric black hole solution to the vacuum Einstein equation, whose metric is given by
\Eq{
ds^2= -f(r)dt^2+ \frac{dr^2}{f(r)} + r^2 d\sigma_n^2;\quad
f(r)= 1- \pfrac{r_0}{r}^{n-1}-\lambda r^2
\label{SSBH:metric}
}
where $d\sigma_n^2$ is a metric of the $n$-dimensional unit sphere $S^n$, and $\lambda$ is a constant related to the cosmological constant by $\lambda= \frac{2\Lambda}{n(n+1)}$. In the asymptotically flat vacuum case with $\lambda=0$, this is the unique regular static black hole solution in four and higher dimensions from the uniqueness theorem.

For this spherically symmetric spacetime, the horizon is a Killing horizon and its location is given by $r=r_h$ in terms of a solution to $f(r_h)=0$. The horizon is obviously homeomorphic to $\RF\times S^n$.

\subsubsection{Kerr black hole}

In the asymptotically flat vacuum case in four dimensions, a regular rotating stationary black hole solution is unique and given by the Kerr solution with the metric
\Eq{
ds^2=
   -\frac{\Delta\rho^2}{\Gamma}dt^2
     +\frac{\Gamma\sin^2\theta}{\rho^2}
     (d\phi-\Omega dt)^2
  +\rho^2\left(\frac{dr^2}{\Delta}+d\theta^2\right),
}
where
\Eq{
\Delta=r^2-2Mr+a^2,\quad
\rho^2=r^2+a^2\cos^2\theta,\quad
\Gamma=(r^2+a^2)^2-a^2\Delta\sin^2\theta,\quad
\Omega=\frac{2aMr}{\Gamma}.
}

The global horizon of this spacetime is again a Killing horizon, and its location is given by the largest solution to $\Delta(r)=0$ as $r=r_h=r_+=M+(M^2-a^2)^{1/2}$. Note that $\Delta(r)=0$ is equivalent to the condition that the Killing orbit spanned by $\pd_t$ and $\pd_\phi$ becomes null. The horizon is topologically $\RF\times S^2$. Because the rotation of the time translation Killing vector $\pd_t$ does not vanish,
\Eq{
\omega=d\inpare{\frac{2aM\cos\theta}{\rho^2}} \neq0,
}
this black hole is rotating. Hence, the time-translation Killing vector is spacelike on the horizon as in Fig.\ref{fig:rotatinghorizon}, and there appears a region called the ergo region where $g_{tt}>0$ as is seen from
$
\rho^2 g_{tt}= a^2\sin^2\theta - \Delta
$.
The existence of this ergo region plays a crucial role in the superradiance instability of a rotating black hole discussed in the next section. The angular velocity of the horizon $\Omega_h$ is determined by the condition that $\pd_t+\Omega_h \pd_\phi$ is a null vector as
\Eq{
\Omega_h = \Omega(r_h)= \frac{2aMr_h}{(r_h^2+a^2)^2}= \frac{a}{2Mr_h}=\frac{a}{r_h^2+a^2}.
}
%

\section{Bound States and Scattering}

\subsection{Particles around a black hole}

\begin{figure}
\begin{minipage}{6cm}
\centerline{
\includegraphics[width=5cm]{\FigDir/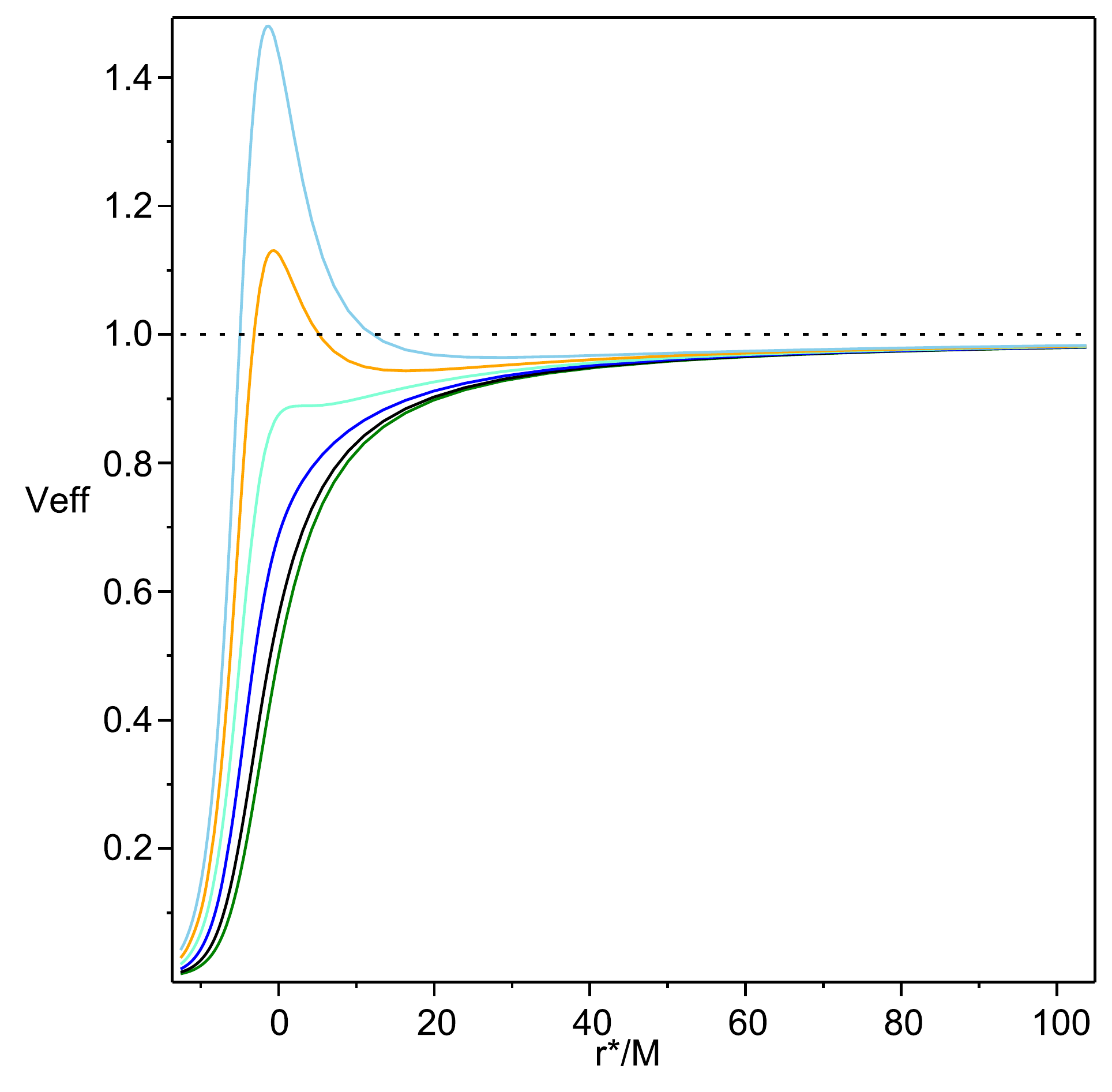}
}
\centerline{\small Massive particle}
\end{minipage}
\begin{minipage}{6cm}
\centerline{
\includegraphics[width=5cm]{\FigDir/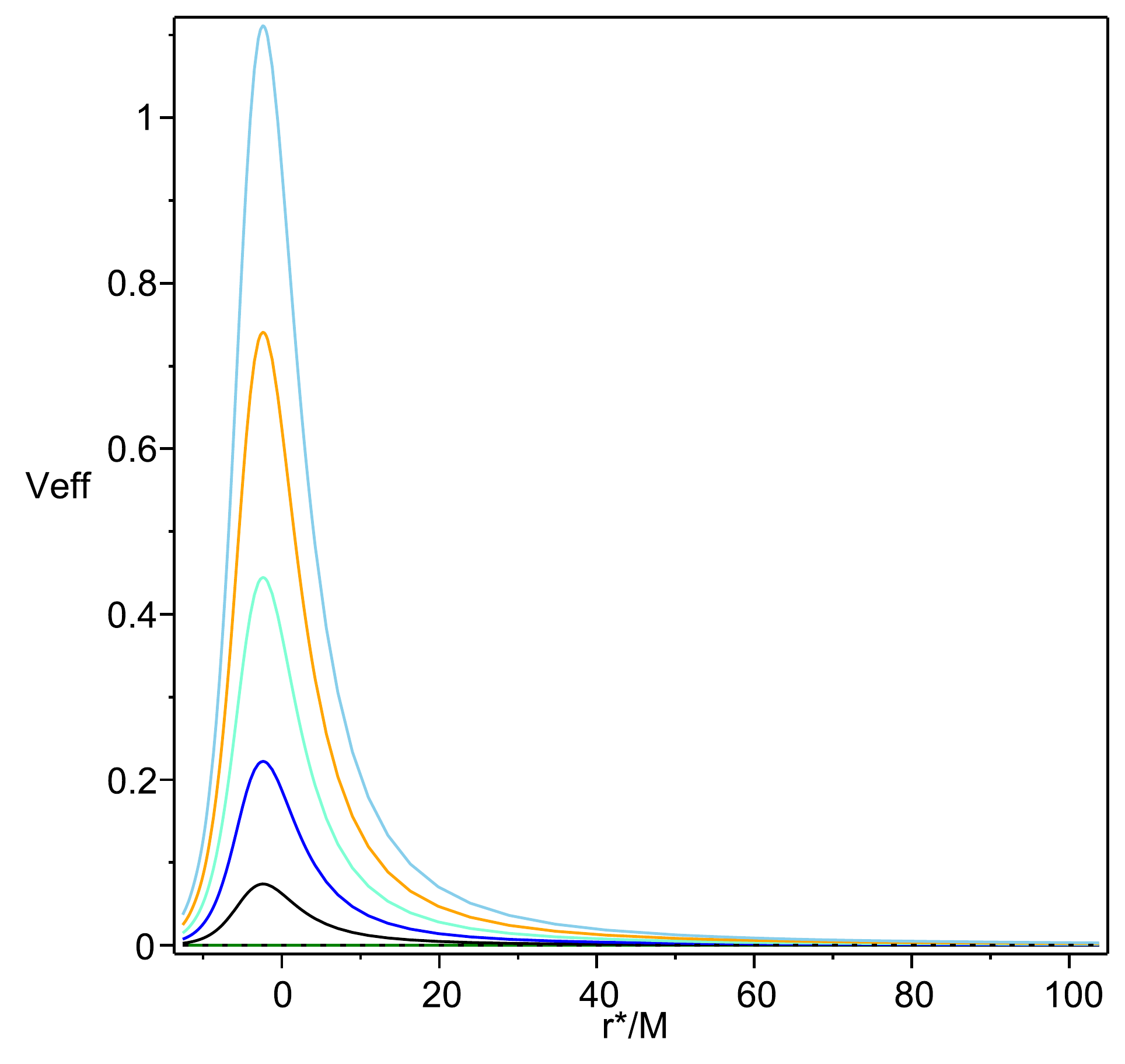}
}
\centerline{\small Massless particle}
\end{minipage}
\caption{The effective potential for a particle with $L=0,\cdots,5$ around the 4D Schwarzschild BH}
\label{fig:PotPeffSSBH}
\end{figure}

\subsubsection{Schwarzschild black hole}

The behavior of a particle around a black hole depends both on the rotation of the black hole and whether the particle is massive or massless.

For example, geodesics around a Schwarzschild black hole with the metric \eqref{SSBH:metric} can be determined by solving the first-order ODE system obtained from the energy and angular-momentum conservation laws,
\Eqrsub{
&& E=-u\cdot\xi = -u_t= f(r)\dot t,\quad
   L = u \cdot\eta = u_\phi = r^2\dot\phi,\\
&& -\epsilon=-f \dot t^2 + \frac{\dot r^2}{f} + r^2\dot\phi^2,
}
where $\epsilon=1$ for a massive particle and $\epsilon=0$ for a massless particle. In particular, their qualitative behaviour can be easily found from the behaviour of the effective potential $V(r) $ in the effective energy conservation law expressed in terms of the $r$ coordinate (see Fig.\ \ref{fig:PotPeffSSBH}):
\Eq{
\dot r^2 + V(r)= E^2;\quad
V(r)= \inpare{\epsilon + \frac{L^2}{r^2}}f(r).
}
In particular, we see that there exist stable bound orbits for a massive particle while there exists no such orbit for a massless particle. Note that in five or higher dimensions, even a massive particle has no stable bound orbit.

\subsubsection{Kerr black hole}

\begin{figure}
\begin{minipage}{6cm}
\centerline{
\includegraphics[width=5cm]{\FigDir/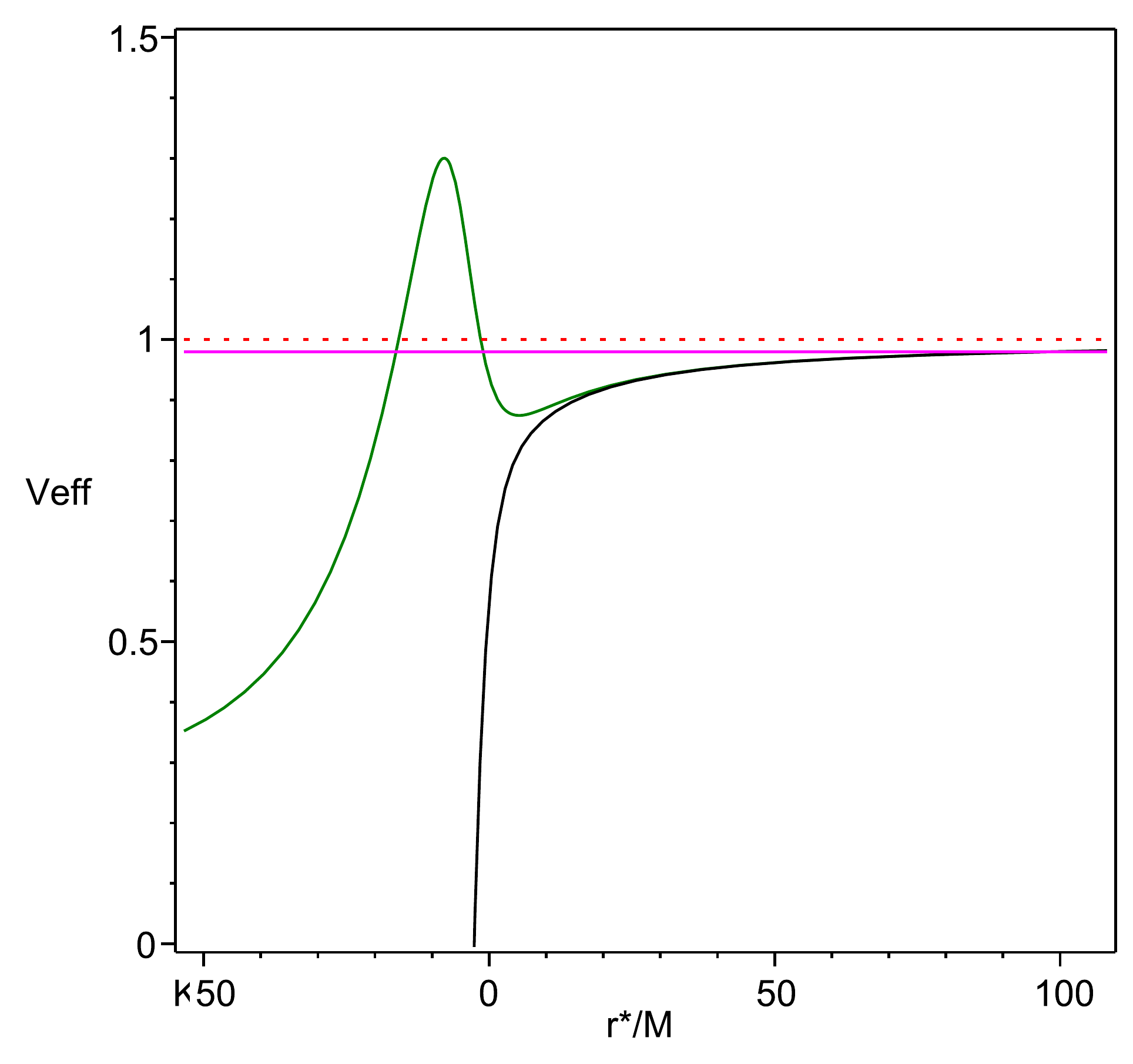}
}
\centerline{\small Massive particle}
\end{minipage}
\begin{minipage}{6cm}
\centerline{
\includegraphics[width=5cm]{\FigDir/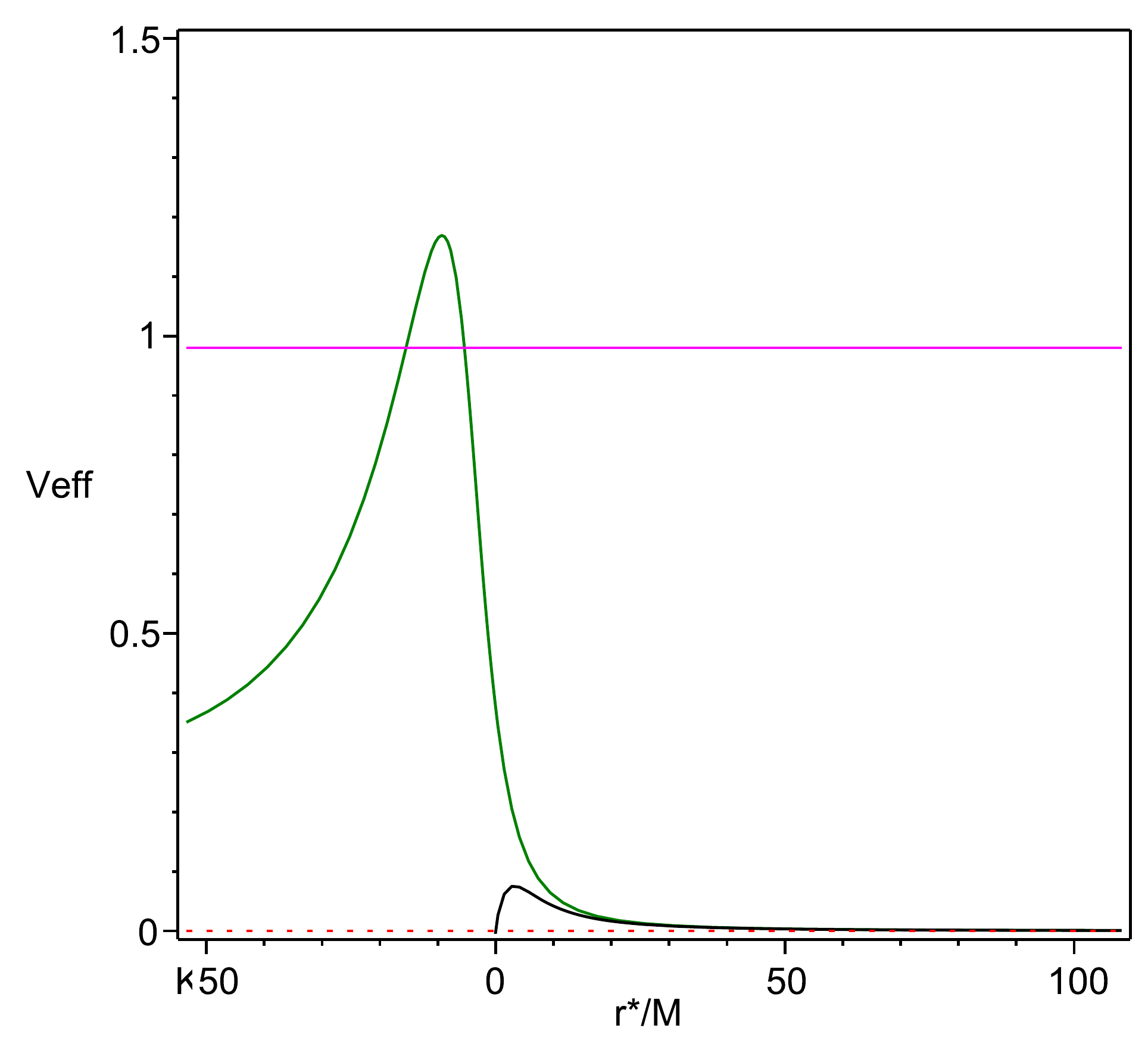}
}
\centerline{\small Massless particle}
\end{minipage}
\caption{The effective potential for particles with $L=3$ (corotating) and $L=-3$ (counter-rotating) around the Kerr BH with $a=0.999$. $r^*$ is the tortoise coordinate defined by $dr^*=(r^2+a^2)dr/\Delta$. }
\label{fig:PotPeffKerrBH}
\end{figure}

For a Kerr black hole, we have to use the Carter constant in addition to the energy and angular momentum to reduce the geodesic equations to a first-order system of ODEs in general. However, orbits on the equatorial plane can be determined by the energy and momentum conservations laws
\Eqrsub{
&& E=-g_{tt}\dot t -g_{t\phi}\dot\phi,\quad
   L =g_{\phi t}\dot t + g_{\phi\phi}\dot\phi,\\
&& -\epsilon=g_{tt}\dot t^2+2g_{t\phi}\dot t \dot \phi + g_{\phi\phi}\dot\phi^2 + \frac{r^2}{\Delta}\dot r^2,
}
or the effective potential for the $r$ coordinate,
\Eq{
\dot r^2 + V(r)= E^2;\quad
V(r)= \frac{\epsilon \Delta}{r^2}-\frac{a^2E^2-L^2}{r^2}
   -\frac{2M(aE-L)^2}{r^3},
}
where the particle is corotating with the black hole for $L>0$ and counter-rotating for $L<0$. As we see from Fig. \ref{fig:PotPeffKerrBH}, the behavior of a particle is largely different for the corotating case and for the counter-rotating case. In particular, the centrifugal force on counter-rotating particles is weaker than that for corotating ones.

\subsection{Massless fields around a Kerr black hole}

\subsubsection{Flux conservation law}

As we saw in the previous subsection, a massless particle incident to a black hole is simply scattered off or absorbed by the black hole, and nothing peculiar happens. However, if we consider the scattering of a massless scalar field, an interesting new phenomenon happens, when the black hole is rotating. This is because a wave behaves as an ensemble of particles.

\begin{figure}
\centerline{
\includegraphics[height=6cm]{\FigDir/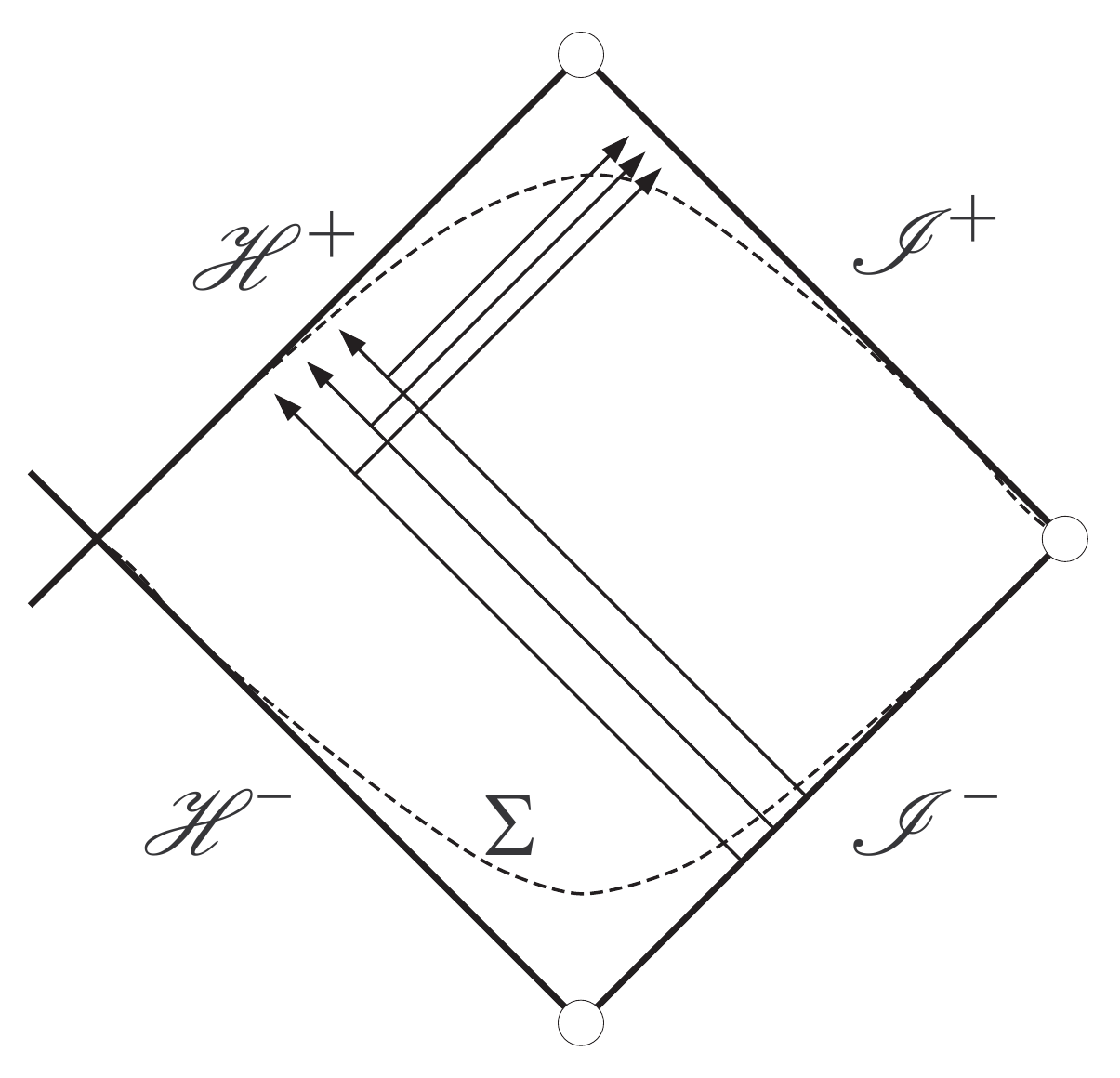}
}
\caption{Scattering of an incidental wave by an AF black hole}
\label{fig:AFBHsuperradiance}
\end{figure}

To see this, let us consider the free Klein-Gordon field around a stationary and axisymmetric rotating black hole satisfying
\Eq{
D^\mu D_\mu \phi=0;\quad D_\mu=\pd_\mu - iq A_\mu,
}
where we have considered a charged black hole with the electromagnetic potential $A_\mu$ and a charged field with charge $q$ for generality. Then, we find that the Klein-Gordon inner product
\Eq{
 N(\phi_1,\phi_2)=- i \int_\Sigma \inpare{\bar\phi_1 D^\mu \phi_2
 - (\bar D^\mu \bar \phi_1) \phi_2 } d\Sigma_\mu
}
does not depend on the choice of a Cauchy surface $\Sigma$ in the DOC. In particular, if we consider the scattering problem as shown in Fig.\ref{fig:AFBHsuperradiance}, we obtain the following relation among the flux coming from infinity $I_{\scri^-}$, the flux absorbed by the black hole $I_{\H^+}$ and the flux escaping to infinity $I_{\scri^+}$ under the assumption that no flux is coming from $\H^-$:
\Eq{
I_{\scri^-}= I_{\scri^+} + I_{\H^+}
\label{FluxConservation:BH:general}
}
%

\subsubsection{Superradiance}

Let us evaluate each of these fluxes. First, because the wave behaves at infinity as
\Eq{
\phi \approx \int d\omega \sum_m \frac{1}{r}\inpare{A^- e^{-i\omega u_-}
  + A^+ e^{+i\omega u_+}} e^{ im\varphi};\quad
u_\pm= t \mp \int dr/f,
}
the fluxes from and to infinity can be expressed as
\Eq{
 I_{\scri^\pm}=
   i \int d{u_\pm}\int_{S^2}d\Omega_2 \lim_{r\tend\infty}
   r^2(\bar\phi\overset{\leftrightarrow}{\pd}_{u_\pm} \phi)
 = \sum_m \int d\omega \omega \langle |A^\pm_{\omega,m}|^2\rangle
_{S^2},
}
where $\EXP{Q}_{S^2}$ is the average of a function $Q$ on $S^2$.

Next, near the future horizon $\H^+$, the wave behaves as
\Eqr{
 \phi &=& \phi(r,\theta) e^{-i\omega t+ im\varphi}
  = \phi(r,\theta)e^{-i\omega_* t + im\tilde\varphi}
\notag\\
 &=& {\phi(r,\theta)e^{i\omega_* r^* }} e^{-i\omega_* v_+ + im\tilde\varphi}
 \approx  C(\theta) e^{-i\omega_* v_+ + im\tilde\varphi},
}
where
$
\omega_*:= \omega - m \Omega_h ,\quad
\tilde\varphi=\varphi -\Omega_h t
$,
and $v_+= t+\int dr (r^2+a^2)/\Delta$.
Because, $v_+$ and $\tilde\varphi$ are regular coordinates around $ \H^+$, this implies that $C$ should be a bounded function of $\theta$. Hence, the flux crossing $\H^+$ can be calculated as
\Eqr{
I_{\H^+} &=& i \int dv_+ \int_{S^2} d^{D-2} \sigma \inpare{\bar\phi (\overset{\leftrightarrow}{\pd}_{v_+}+2iq\Phi)\phi}_{\H^+}
\notag\\
&=& \sum_m \int d\omega { (\omega_*-q\Phi_h) } (r_h^2+a^2)
 \langle|C_{\omega,m}|^2\rangle _{S^2}.
}

Inserting these expressions for fluxes into \eqref{FluxConservation:BH:general}, we obtain
\Eq{
\omega \langle |A^-_{\omega,m}|^2\rangle
   =\omega \langle |A^+_{\omega,m}|^2\rangle
    +\inpare{\omega- m \Omega_h - q\Phi_h} (r_h^2+a^2)\langle |C_{\omega,m}|^2\rangle.
}
If we define the transmission rate $T$ and the reflection rate $R$ of the wave by the black hole as
\Eq{
T:= I_{\H^+}/ I_{\scri^-},\quad
R:=I_{\scri^+}/ I_{\scri^-},
}
we obtain $R>1$ because $T+R=1$ and $T<0$ when the condition
\Eq{
\omega_*-q\Phi_h =\omega- m\Omega_h - q \Phi_h < 0
\label{SuperradianceCondition}
}
 is satisfied, where $\Phi_h$ is the electric potential of the black hole.  That is, the scattered wave is amplified and  have a larger flux than the incoming wave. This phenomenon is called the {\em superradiance} and the condition \eqref{SuperradianceCondition} is called the superradiance condition.

\begin{figure}
\centerline{
\includegraphics[height=6cm]{\FigDir/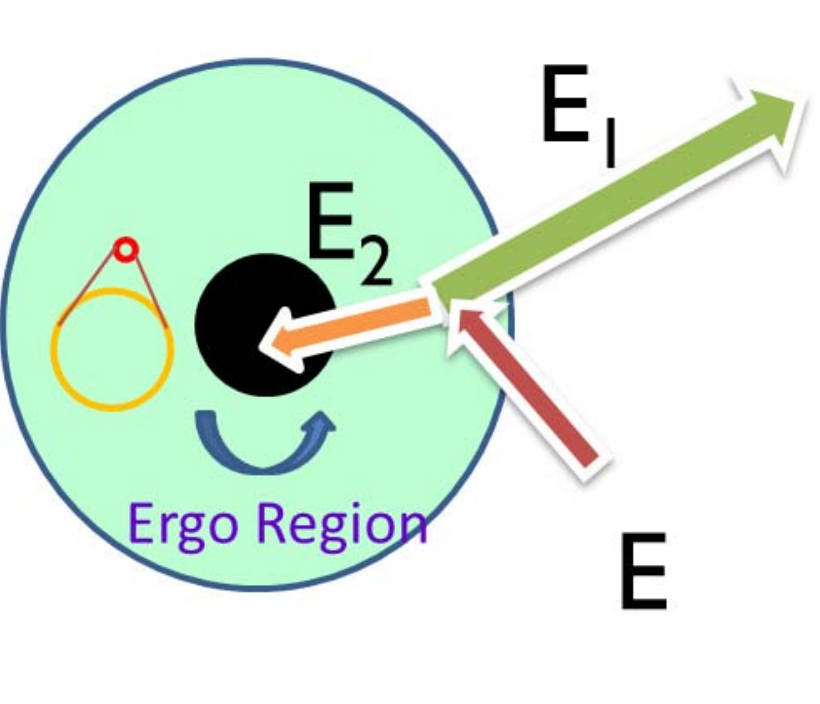}
}
\caption{The Penrose process in the ergo region}
\label{fig:PenroseProcess}
\end{figure}

\subsubsection{Penrose process}

This curious phenomenon is closely related to the Penrose process\cite{Penrose.R&Floyd1971} in the ergo region. In this region, because the time-translation Killing vector $\xi$ is spacelike, the energy $E=-p\cdot\xi$ with respect to the spatial infinity can become negative even for a future-directed timelike 4-momentum $p^\mu$. Hence, if a particle incidental to the ergo region decays into two particles there, the outgoing particle can have a large energy than the incidental one if the other decay product has a negative energy as shown in Fig.\ref{fig:PenroseProcess}. This is called the {\em Penrose process}. The produced negative energy particle cannot go outside of the ergo region, and is eventually absorbed by the black hole to reduce its mass. Hence, the Penrose process makes it possible to extract energy (and angular momentum) from a rotating black hole. The superradiance above can be regarded as being produced by the same mechanism because the superradiance condition can be written
\Eq{
k\cdot p>0,\quad p_\mu \phi=(-i\pd_\mu - q A_\mu)\phi,
}
where $k=\pd_t + \Omega_h \pd_\varphi$ is the future-directed null generator of the horizon $\H^+$. This equation implies that $p$ is a past-directed timelike vector because the physical momentum $p$ is timelike.

\section{Superradiance Instability}

Superradiation does not cause any physical problem by itself. However, if we consider the gedanken experiment to put a rotating black hole inside a box with reflective boundary, superradiance provokes an instability called a black hole bomb\cite{Zeldovich.Y1971,Press.W&Teukolsky1972,Cardoso.V&&2004} because amplified  waves of a massless field by superradiance are reflected back toward the black hole by the surrounding mirror wall of the box, and the repetition of this process produces an unbounded growth of the waves. This instability lasts until the central black hole loses its whole angular momentum.

In reality, of course, it is impossible to put a black hole in a box, but Damour, Deruelle and Ruffini\cite{Damour.T&Deruelle&Ruffini1976} pointed out that we can realize such a situation practically just by considering a massive field instead of a massless field around a rotating black hole. This is because a massive field can have bound states and the effective potential provide an effective outer wall for the field as shown in Fig.\ref{fig:EffPotKerrBHKG}. In this section, we look at the instability of such a massive scalar around a Kerr black hole in details.


\subsection{Massive scalar equation around a Kerr black hole}

\begin{figure}
\begin{minipage}{6.5cm}
\centerline{
\includegraphics[width=6cm]{\FigDir/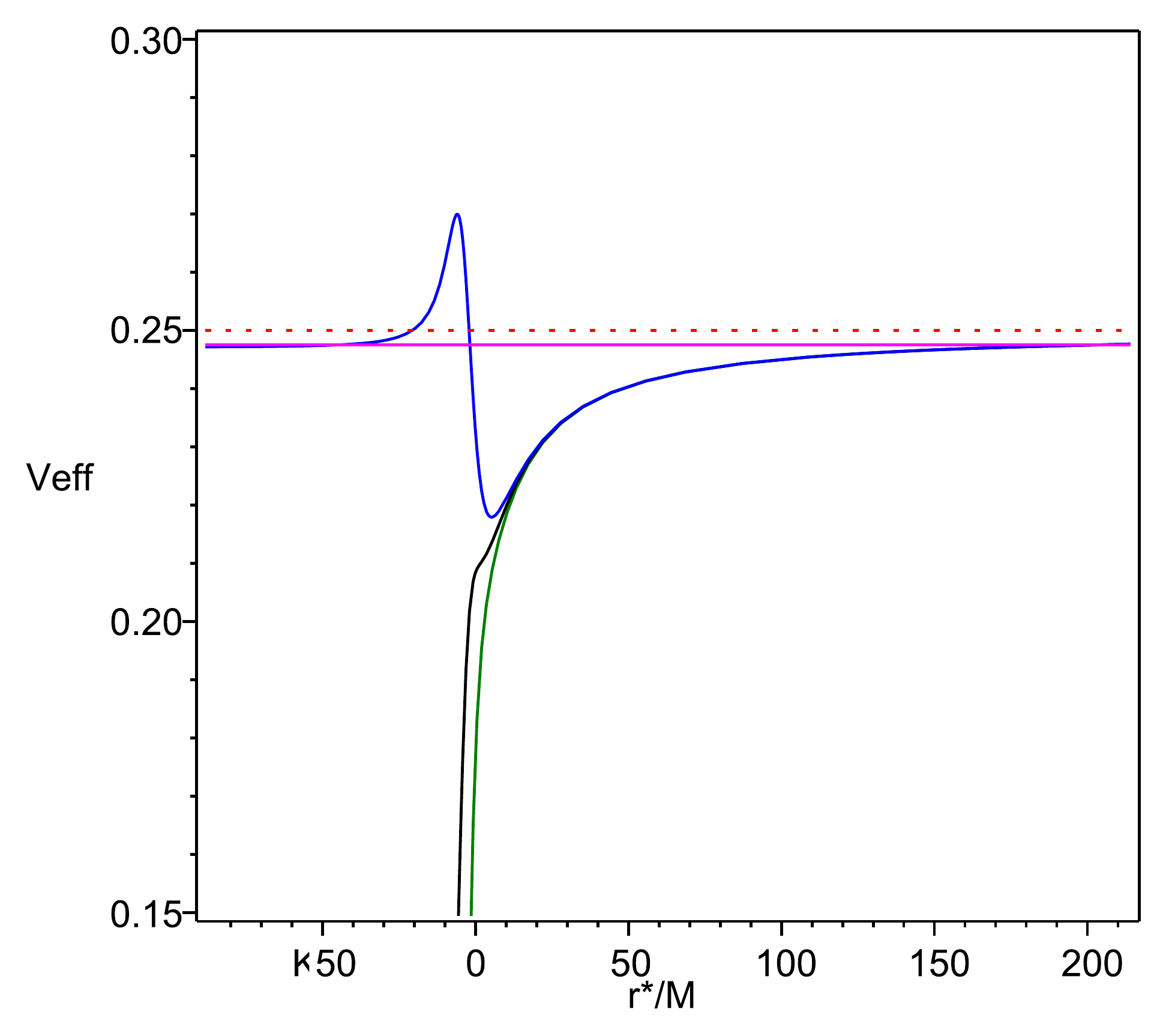}
}
\centerline{\small $\mu M=0.5$, $l=1$}
\end{minipage}
\begin{minipage}{6cm}
\centerline{
\includegraphics[width=6cm]{\FigDir/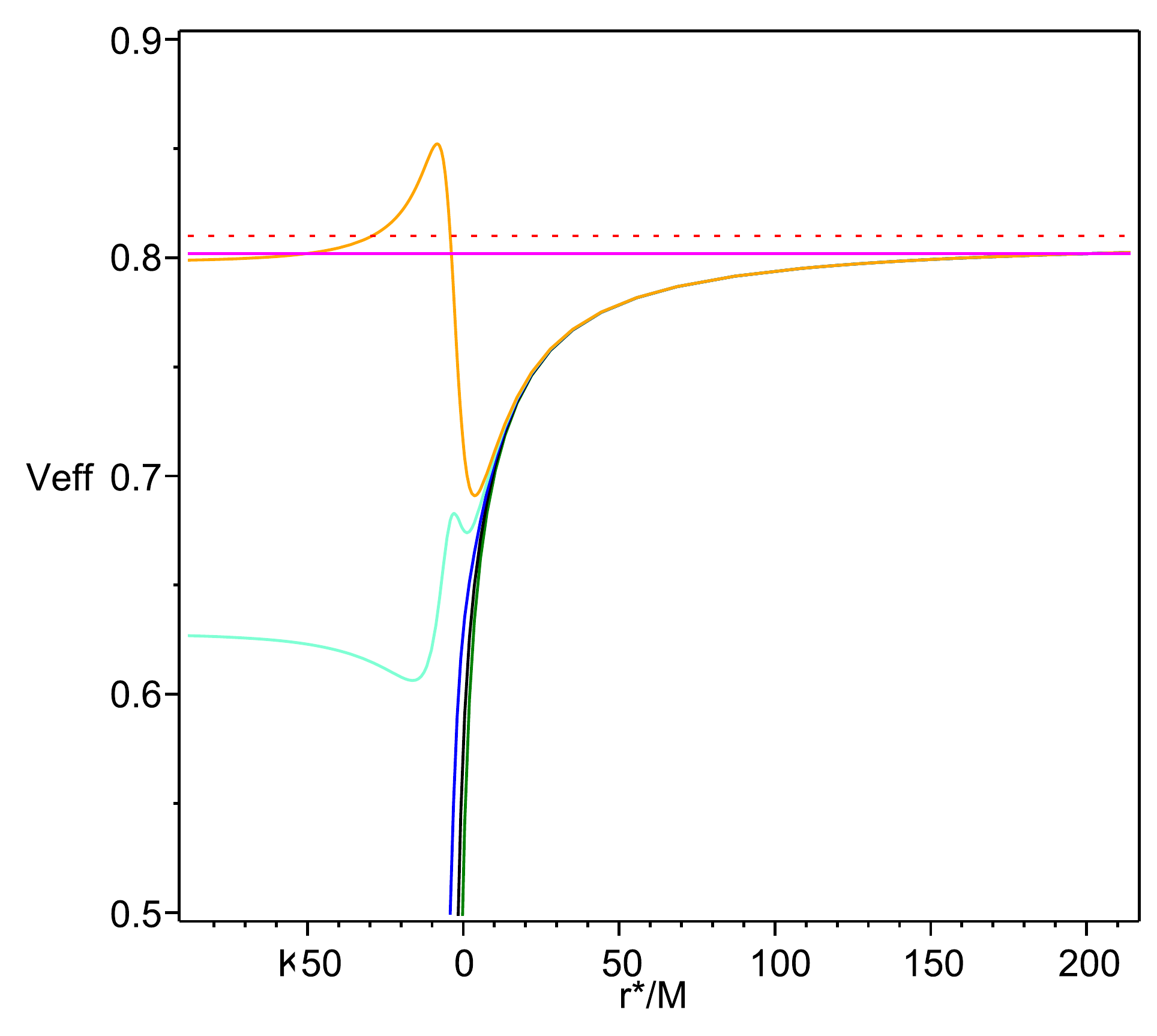}
}
\centerline{\small $\mu M=0.9$, $l=2$}
\end{minipage}
\caption{The effective potential for a massive scalar field around the Kerr BH with $a=0.999$.}
\label{fig:EffPotKerrBHKG}
\end{figure}

The field equation for a massive free scalar field
\Eq{
(\Box -\mu^2)\Phi=0
}
can be separated by
\Eq{
\Phi = R_{lm}(r)S_{lm}(\theta) \exp(-i\omega t + im \phi),
\label{Phi:separated}
}
into the ODE for  the angular mode function $S_{lm}(\theta)$,
\Eq{
\frac{1}{\sin\theta}\frac{d}{d\theta}\inpare{\sin\theta\frac{d S_{lm}}{d\theta}}
 + \insbra{a^2(\omega^2-\mu^2)\cos^2\theta -\frac{m^2}{\sin^2\theta}
 + \Lambda_{lm}} S_{lm}=0,
\label{ODE:angular}}
with the separation constant $\Lambda_{lm}$, and the ODE for the radial mode function $R_{lm}(r)$,
\Eqr{
\frac{d}{dr}\inpare{\Delta \frac{dR_{lm}}{dr}}
 + \Big[&&\frac{\omega^2(r^2+a^2)^2-4Mam\omega r + m^2a^2}{\Delta}
\notag\\
 &&- (\omega^2a^2+\mu^2r^2 + \Lambda_{lm}) \Big]R_{lm}=0.
\label{ODE:radial}
}

The equation \eqref{ODE:angular} for the angular mode function, as an ODE with respect to the independent variable $x=\cos\theta$, have two regular singularities at $x=\pm1$ with index $\pm m/2$ and an irregular singularity at infinity. Hence, it has a series of regular solutions $S_{lm}=S^m_l(x;c)$ with $c=a(\omega^2-\mu^2)^{1/2}$ on the interval $-1\le x\le 1$ corresponding to discrete values of $\Lambda_{lm}$ labeled by an integer $l=0,1,2,\cdots$.  In the limit $c\tend0$, it reduces to the standard associated Legendre function as
\Eq{
S^m_l \tend P^m_l,\quad
\Lambda_{lm}\tend l(l+1).
}

The equation \eqref{ODE:radial} for the radial mode function can be rewritten in terms of $u=(r^2+a^2)^{1/2}R_{lm}$ as
\Eq{
\frac{d^2u}{dr^*{}^2}+ \insbra{\omega^2-V(r,\omega)} u=0,
}
which defines the effective potential
\Eqr{
V &=& \frac{ \mu^2\Delta}{r^2+a^2}
 + \frac{4am\omega Mr-a^2m^2+\Delta[\Lambda_{lm}+(\omega^2-\mu^2)a^2]}
  {(r^2+a^2)^2}
\notag\\
&& + \frac{\Delta (2Mr^3+a^2r^2-4Ma^2 r + a^4)}{(r^2+a^2)^4}.
}
This potential always approaches $\mu^2$ at $r=\infty$ and $\omega^2-\omega_*^2$ at horizon as illustrated in Fig. \ref{fig:EffPotKerrBHKG}. In order to study the superradiance instability, we have to look for a unstable solution to this equation satisfying the following boundary condition:
\Eqrsub{
\text{At infinity} &:&
R_{lm} \sim  \frac{B_{lm}}{r} e^{+i k r^*},\quad
k=(\omega^2-\mu^2)^{1/2}.
\label{BC:infinity}
\\
\text{At horizon} &:&
R_{lm}\sim C_{lm} e^{-i\omega_* v_+ + im\tilde \phi}
 \sim C_{lm} e^{-i\omega_* r^*}e^{-i\omega t + im\phi}.
\label{BC:horizon}
}
Note that the first condition implies that no wave is coming from infinity and the second condition requires that waves are purely infalling at the horizon.

\subsection{Instability condition}

From the field equation, for the solution of the form \eqref{Phi:separated}, we obtain the following energy integral:
\Eqr{
0&=& i\int \frac{d\phi}{2\pi} \int d\theta \sin\theta
 \int dr \rho^2 (\pd_t\Phi)^*(\Box-\mu^2)\Phi
 \notag\\
 &=& \insbra{ \int  d\theta \sin\theta
   (-\omega^*)\Delta \Phi^* \pd_r\Phi}_{r=r_h}^{r=r_\infty}
   \notag\\
  && + \int dr \int  d\theta \sin\theta
   \Big[ \rho^2 |\omega|^2( \omega g^{tt}-2 m g^{t\phi}) |\Phi|^2
   \notag\\
  &&\qquad
   +\omega^*\rho^2\inpare{m^2g^{\phi\phi}   + \mu^2 } |\Phi|^2
       +\omega^* \inpare{\Delta|\pd_r\Phi|^2 + |\pd_\theta \Phi|^2 }
     \Big].
}
If the solution satisfies the boundary condition \eqref{BC:infinity} and \eqref{BC:horizon}, this identity can be written in terms of $\omega=\omega_R  + i \omega_I$ and $\tilde R=R_{lm}\exp(i\omega_* r^*)$ as
\Eq{
B \omega_R(m\Omega_h-\omega_R)+\omega_I^2 (C_1 +C_2\omega_I) = A \omega_I,
\label{EnergyIntegralId}
}
where
\Eqrn{
&& B= (r_h^2+a^2) e^{2\omega_I v_+} |\tilde R|^2_{r=r_h},\\
&& C_1=\int dr 2r e^{2\omega_I v_+}|\tilde R|^2,\\
&& C_2=a^2 \int dr e^{2\omega_I v_+} \int d\theta\sin^3\theta |\tilde R S^m_l|^2,
}
and
\Eqr{
&& A = \int dr e^{2\omega_I v_+} \int d\theta \sin\theta
 \Big[   |\tilde R|^2|\pd_\theta S^m_l|^2
 + \rho^2{ (-g_{tt})} |\pd_r \tilde R|^2 |S^m_l|^2
\notag\\
&& \quad + a^2 \left|\sin\theta\pd_r\tilde R
 - i\omega_R^* \frac{(r^2+a^2)}{a^2\sin\theta}\tilde R\right|^2  |S^m_l|^2
 +\inrbra{
  \frac{2Q^2}{\sin^2\theta} + P \omega_R^2 + \mu^2\rho^2 } |\tilde R S^m_l|^2
  \Big],
}
where $Q$ is a real quantity and $P$ is
\Eqr{
&& P = \rho^2(r^2+a^2)^2
 \inrbra{\rho^2  + 4a^2\sin^2\theta}(g_{tt})^2
 \notag\\
&& \quad
+ 8Ma^2\sin^2\theta
  \insbra{ r(r^2+a^2) {(- g_{tt})}
  + \frac{a^2Mr^2 \sin^2\theta}{\rho^2}}
  \inpare{\rho^2+ 2a^2\sin^2\theta } .
}

Obviously, $A, B, C_1$ and $C_2$ are all positive and finite if $\omega^2<\mu^2$, which guarantees that the wavefunction falls off exponentially at infinity, and if the contribution from the integral in the ergo region where $(-g_{tt})<0$  is not dominant. In that case, from the identity \eqref{EnergyIntegralId}, it follows that $\omega_I>0$ when the superradiance condition is satisfied. This implies that the scalar field grows exponentially in time as $\Phi\propto \exp(\omega_I t)$. If any of these conditions is not satisfied, we cannot draw a definite conclusion about the instability. Thus, we obtain the following sufficient conditions for the occurrence of superradiant instability\cite{Zouros.T&Eardley1979}:
\Blist{}{}
\item[i)]  The mode is bounded.
\item[ii)]  The wavefunction is peaked far outside the ergo region.
\item[iii)] $\omega$ is nearly real: $|\omega_I |\ll \omega_R$.
\item[iv)] $\omega$ satisfies the superradiance condition:  $\omega_R <  m\Omega_h$.
\Elist

\subsection{Growth rate}

In this section, we estimate the growth rate of the superradiance instability of a massive scalar field with mass $\mu$ around a Kerr black hole with mass $M$ by three different methods. The first is based on the WKB approximation that is valid when $M \mu\gg1$ in the absolute units $c=\hbar=G=1$. The second is the matched asymptotic expansion method that is valid when $M \mu \ll1$. Unlike these semi-analytic methods, the third is a purely numerical method based on the continued fraction that was first introduced by Leaver\cite{Leaver.E1985}. In this section, we adopt the absolute units unless otherwise stated.

\subsubsection{Large mass case}

\begin{figure}
\centerline{
\includegraphics[height=6cm]{\FigDir/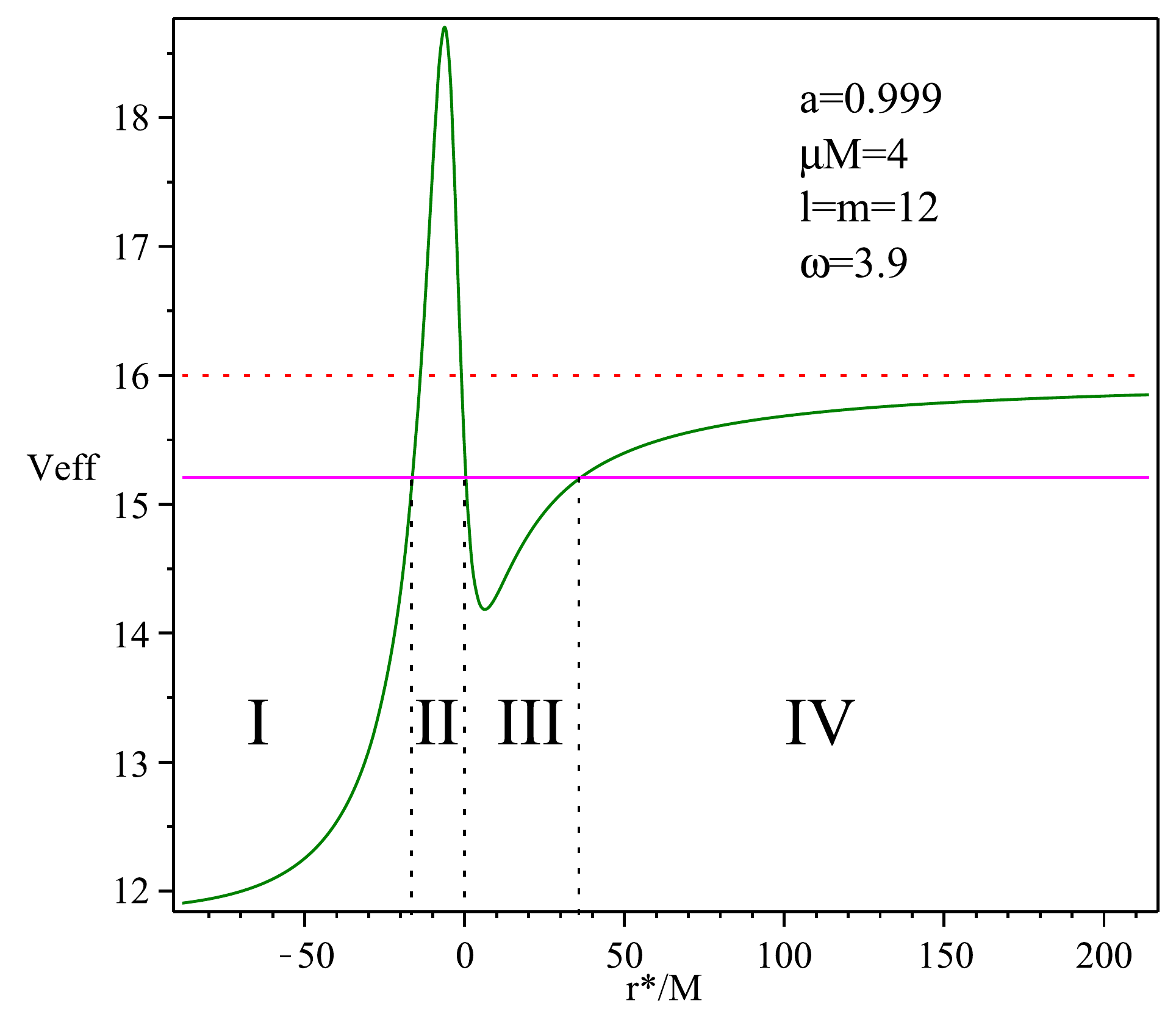}
}
\caption{Division into four regions for the WKB approximation}
\label{fig:PotWKB}
\end{figure}

In the absolute units, the typical scale of the background geometry is $M$, while the Compton wavelength of a particle with mass $\mu$ is $1/\mu$. Hence, when $M\mu \gg1$, the wave length of the scalar field is much shorter than the geometrical scale, and the WKB solution provides a good approximation for the wavefunction. The growth rate of the superradiance instability with this approximation was first estimated by T.J.M. Zouros and D.M. Eardley\cite{Zouros.T&Eardley1979}. We follow their arguments but some numerical error in their result is corrected.

Because we are interested in the superradiant instability, we consider the quasi-bound state with $\omega^2<\mu^2$. Because the instability rate becomes larger as the wave amplitude at horizon is larger, we only consider the case $\omega_R\simeq \mu$ in which the tunneling transition late through the potential barrier becomes maximum. For such a value of $\omega_R$, the range of the coordinate $r$ is divided into four regions I($r<r_1$), II($r_1<r<r_2$), III($r_2<r<r_3$) and IV($r>r_3$) according to the sign of $\omega^2-V(r)$ as shown in Fig.\ref{fig:PotWKB}. In the oscillatory regions I and III where $\omega^2> V(r)$, the WKB solution for the radial mode function $u=(r^2+a^2)^{1/2}R_{lm}$ can be written in the form
\Eq{
u=  k(r^*)^{-1/2} \inrbra{A_+ e^{i\Theta(r)}+ A_- e^{-i\Theta(r)}},
}
where $A_\pm$ are constants and
\Eq{
\Theta(r)=\int^{r^*}_{r^*_0} k(u) du,\quad k(r^*)=(\omega^2-V(r))^{1/2}.
}
Here, $r^*_0$ is the $r^*$ coordinate of a reflection point where $\omega^2=V(r)$. Meanwhile, in the regions II and IV where $\omega^2<V(r)$, the WKB solution reads
\Eq{
u= \kappa(r^*)^{-1/2} \inrbra{ B_- e^{-I(r)}+ B_+ e^{I(r)}},
}
where $B_\pm$ are constants and
\Eq{
I(r)=\int^{r^*}_{r^*_0} \kappa(u) du,\quad \kappa(r^*)=(V(r)-\omega^2)^{1/2}.
}

Because these WKB solutions diverge at the reflection points where $k$ and $\kappa$ vanish, we cannot directly connect these WKB solutions in different regions by the standard regularity requirement. The standard method to cope with this situation is to utilize the exact solution around each reflection in the case where the potential is locally approximated by a linear function of $r^*$, which can be written in terms of Airy functions. In the oscillatory region, it can be written in terms of the Bessel functions $J_{\pm1/3}$ as
\Eqr{
u &=&
  \sqrt{\frac{|\Theta|}{k}}\inrbra{C_+ J_{1/3}(|\Theta|)+ C_- J_{-1/3}(|\Theta|)}
  \notag\\
  &\sim & \frac{1}{\sqrt{2\pi k}}
   \insbra{\inpare{C_+ e^{-\frac{5\pi i}{12}}+C_- e^{-\frac{\pi i}{12}}}e^{i|\Theta|}
   +\inpare{C_+ e^{\frac{5\pi i}{12}}+C_- e^{\frac{\pi i}{12}}}e^{-i|\Theta|} }
}
and in the tunneling region, as
\Eqr{
u &=&
  \sqrt{\frac{|I|}{\kappa}}\inrbra{-C_+ I_{1/3}(|I|)+ C_- I_{-1/3}(|I|)}
  \notag\\
 &\sim& \frac{1}{\sqrt{2\pi \kappa}}
  \insbra{\inpare{C_- -C_+} e^{|I|}
   + \inpare{C_-e^{-\frac{\pi i}{6}} - C_+ e^{-\frac{5\pi i}{6}}} e^{-|I|} }.
}.

By applying this method to connecting adjacent regions starting from the region I with the infalling boundary condition $A^I_+=0$, we obtain the following relations in order:
\Eqrsub{
&& A^{\rm I}_-= A_0,\quad A^{\rm I}_+=0,\\
&& B^{\rm II}_+= e^{-\pi i/4}A_0,\quad B^{\rm II}_-=0,\\
&& A^{\rm III}_+ =-i A^{\rm III}_-=-i e^{I_{\rm II}} A_0,\\
&& B^{\rm IV}_-= e^{-3\pi i/4}e^{I_{\rm II}+ i\Theta_{\rm III}} A_0,
}
and
\Eq{
(e^{\pi i/3}-1) B^{\rm IV}_+= 2 e^{5\pi i/12} e^{I_{\rm II}}\cos\Theta_{\rm III},
}
where
\Eq{
I_{\rm II}=\int_{r^*_1}^{r^*_2} \kappa(r)dr^*,\quad
\Theta_{\rm III}=\int_{r^*_2}^{r^*_3} k(r)dr^*.
}
Here, we have taken the base point for each phase integral $I$ or $\Theta$ at the left end of each  region except for the region I for which $r=r_1$ is taken. Because we are considering a bound mode with $\omega_R^2<\mu^2$, $B^{\rm IV}_+$ should vanish. This leads to the Bohr-Sommerfeld quantization condition for the bound state frequency,
\Eq{
\omega=\omega_n:\quad
\int_{r^*_2}^{r^*_3} k(r)dr^*=\inpare{n+ \frac{1}{2}}\pi,\quad n=0,1,\cdots.
}
%

\begin{figure}
\centerline{
\includegraphics[height=7cm]{\FigDir/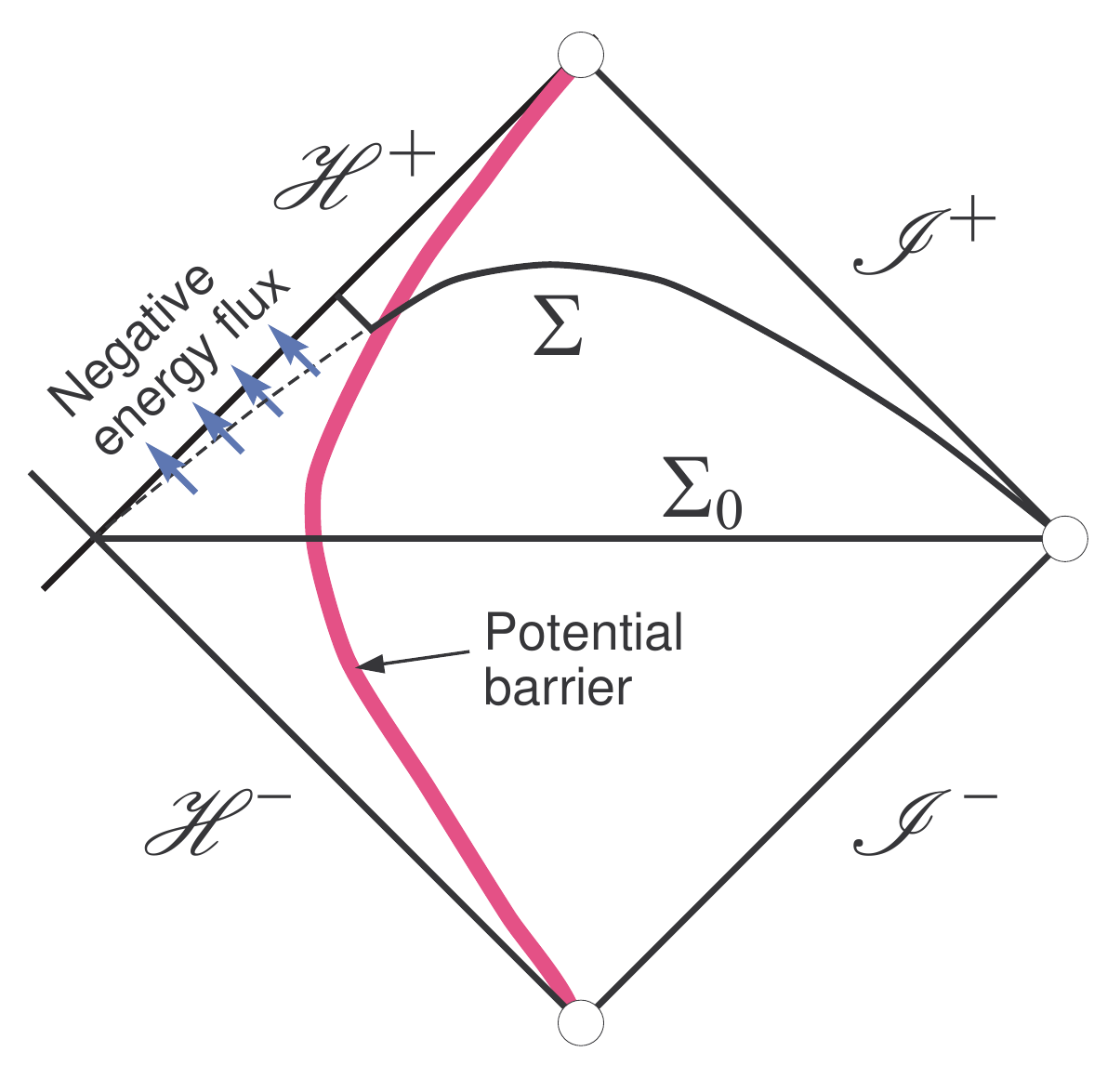}
}
\caption{Flux integral to estimate the growth rate}
\label{fig:AFBH_SRfluxint}
\end{figure}

Now, we can estimate the instability growth rate using this WKB approximation for a quasi-bound state solution. We first note that the KG inner product for an exact complex solution ($\propto \exp(-i\omega t)$) representing an unstable mode has to vanish if it is estimated on a hypersurface $\Sigma_0$ with the inner boundary at the bifurcating 2-sphere (see Fig.\ref{fig:AFBH_SRfluxint}). It is because the boundary of the hypersurface does not move by a time translation and as a consequence, the KG inner product must be time-independent, while it has to be proportional to $\exp(2\omega_I t)$. This implies that the integrand is negative in a region close to the horizon on such a hypersurface. This can be confirmed by look at the explicit expression for the KG inner product,
\Eq{
N(\Phi,\Phi)=e^{2\omega_I t}\int_{r_h}^\infty \inrbra{
 \omega_R \inpare{(r^2+a^2)^2- a^2\Delta \zeta}  -2maMr}
 |R_{lm}|^2\frac{dr}{\Delta},
}
where $\zeta$ is the average $\EXP{\sin^2\theta |S^m_l|^2}$ over $S^2$.

Thus, this type of a hypersurface cannot be used to estimate the instability growth rate. We therefore take a hypersurface $\Sigma$ that has the inner boundary at the future horizon and is parallel to the infalling null direction inside the potential barrier as shown in Fig. \ref{fig:AFBH_SRfluxint}. Then, because the wave is almost purely infalling inside the potential barrier due to the boundary condition at the horizon, the flux integral on this part of the hypersurface $\Sigma$ can be neglected. Further, we can confirm that the peak position of the barrier is around the boundary of the region where the flux in the $\nabla t$ direction is negative. Hence, by translating this surface by time translation, we obtain from the flux conservation
\Eq{
-\omega_* (r_h^2+a^2)|\tilde R|^2 = 2\omega_I N_{\Sigma'}(\Phi,\Phi).
}
From this, $\omega_I$ can be estimated as
\Eqr{
\omega_I &=& \frac{1}{2}\gamma e^{-2 I_{\rm II}},
\label{SRinstability:growthrate:WKB}
\\
\gamma^{-1} &\simeq& \int_{r_2^*}^{r_3^*} \frac{dr^*}{k(r)}
 4\cos^2\inpare{\Theta-\frac{\pi}{4}} \inrbra{\omega_n\inpare{1-\frac{a^2\zeta \Delta}{(r^2+a^2)^2}}-\frac{2maMr}{(r^2+a^2)^2}} .
}

Zouros and Eardley numerically estimated the instability growth rate for wide ranges of the parameters $a/M,\mu M, l ,m$ and $\omega$ using this formula. They found that the growth rate becomes the largest for i) the smallest $l $, ii) the largest possible $m$, i.e., $m=l $, iii) the largest possible $a/M$, i.e., $a/M\simeq1$,  and iv) the largest possible $\omega_R$, i.e., $\omega_R\sim 0.98\mu < m\Omega_h$, and gave the estimate for  the maximum growth rage for a given $\mu M$,
\Eq{
M\omega_I \sim 10^{-7} \exp\inpare{-1.84 \mu M},
\label{SRinstability:growthrate:largemass}
}
The very tiny prefactor $10^{-7}$ comes from $\gamma$ in \eqref{SRinstability:growthrate:WKB}.

\subsubsection{Small mass case}

In the limit $\mu M\ll1$, we can estimate the instability growth rate by a different method\cite{Detweiler.S1980}(Cf. \cite{Rosa.J2010}). This method utilizes the fact that we can approximate the mode function $R_{lm}$ by known analytic functions in two asymptotic regions.

First, in the asymptotic region $r\gg M$, we can approximate the ODE \eqref{ODE:radial} for $R=R_{lm}(r)$ by
\Eq{
\frac{d^2(rR)}{dr^2} + \inpare{\omega^2-\mu^2 + \frac{2M\mu^2}{r}-\frac{l(l+1)}{r^2}}(rR)\approx0,
}
which has exactly the same form as that of the Schr\"odinger equation for a hydrogen atom. Hence, when $\sigma^2=\mu^2-\omega^2>0$, it has a sequence of quasi-bound state solutions,
\Eqr{
&&R = \frac{A}{x} W_{\nu,l+1/2}(x) \sim e^{-x/2}x^\nu \quad(x=2\sigma r\gg1),\\
&& \nu = M\mu^2/\sigma = l+n+1 + \delta \nu,\quad (n=0,1,2,\cdots),
}
where $\delta \nu$ is a complex number representing the deviation from the exactly hydrogen-type wavefunction.  This solution behaves in the region  $\sigma M \ll x \ll 1$  as
\Eq{
R \approx A (-1)^n \frac{(2l+1+n)!}{(2l+1)!} x^{l}
    +A (-1)^{n+1} n!(2l)!{\delta \nu }x^{-l-1}.
\label{R:expansion:far}
}

Next, in the region $\mu r\ll l $, the ODE \eqref{ODE:radial} can be approximated by
\Eq{
z(z+1)\frac{d}{dz}\insbra{z(z+1)\frac{dR}{dz}}
 + \inrbra{P^2-l(l+1)z(z+1)} R=0,
}
where
\Eq{
z=\frac{r-r_+}{r_+-r_-},\quad
P=-\frac{2Mr_+}{r_+-r_-} \omega_*
}
This can be exactly solved, and the solution infalling into horizon is given by
\Eq{
R=C \pfrac{z}{z+1}^{iP} F(-l,l+1,1+2iP;-z).
}
This solution behaves in the overlapping region $1 \ll z \ll l/(\omega_R M)$ as
\Eq{
R \approx C\frac{(2l)!\Gamma(1+2iP)}{l! \Gamma(l+1+2iP)} z^l
 + C (-1)^{l+1} \frac{l! \Gamma(1+2iP)}{(2l+1)!\Gamma(- l+2iP)} z^{-l-1}.
}
Matching the two approximate solutions in the overlapping region,  we obtain
\Eq{
\delta\nu = 2iP\insbra{2\sigma(r_+-r_-)}^{2l+1}\frac{(2l+1+n)!}{n!}
 \insbra{\frac{l!}{(2l)!(2l+1)!}}^2 \prod^l_{j=1}(j^2+4P^2).
}
This determines the instability growth rate as
\Eqrsub{
\omega_R  &\simeq& \mu\inrbra{1 -\pfrac{\mu M}{l+1+n}^2}^{1/2}\approx \mu
\label{hydrogenlevel}
\\
\omega_I  &=& 2 \gamma  \mu r_+ (m\Omega_h-\mu) (\mu M)^{4l+4},
}
where
\Eq{
\gamma = \frac{2^{4l+2}(2l+1+n)!}{n!(l+1+n)^{2l+4}}
  \pfrac{l!}{(2l)!(2l+1)!}^2
  \prod^l_{j=1} \insbra{ j^2\inpare{1-a^2/M^2}
 + 4r_+^2 (\mu-m\Omega_h)^2}.
}
Clearly, for $l =m=1$ and $a/M\sim1$, the growth rate takes the maximum value
\Eq{
\omega_I \approx \frac{a}{24M^2} (\mu M)^9.
\label{SRinstability:growthrate:smallmass}
}
%

\begin{figure}
\includegraphics[height=7cm]{\FigDir/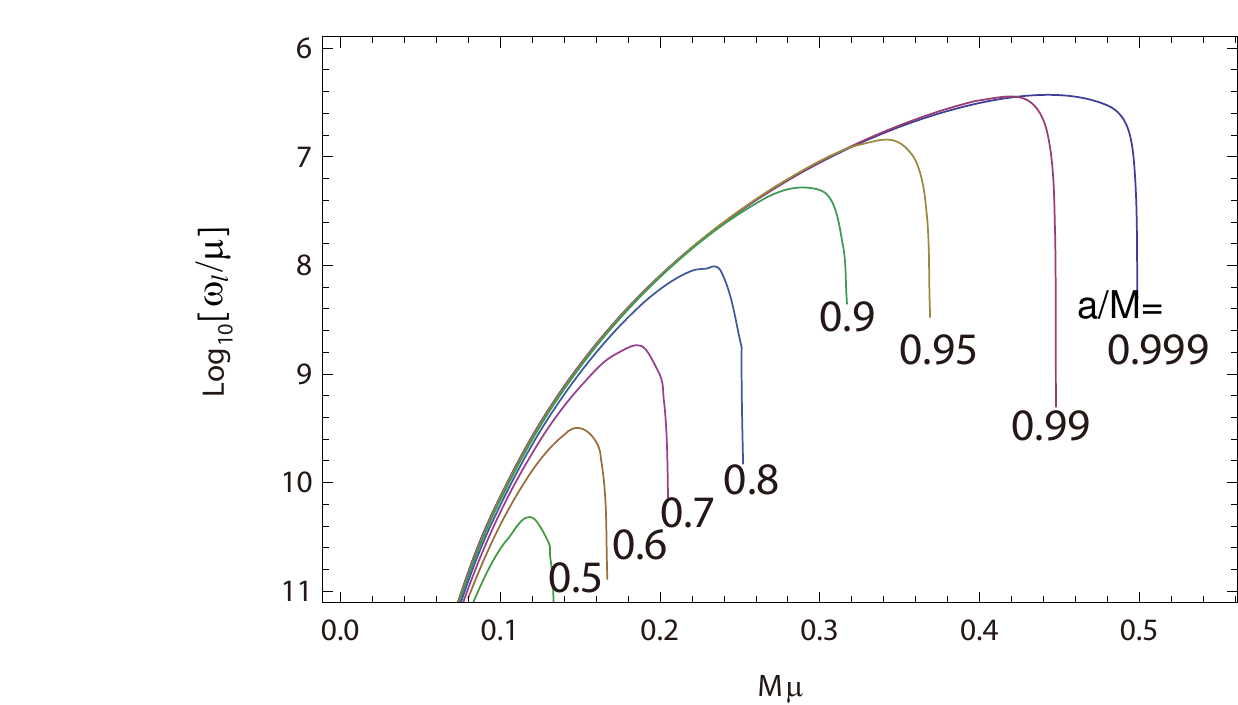}
\caption{The instability growth rate for $l =m=1$.}
\label{fig:GrowthRate}
\end{figure}

\subsubsection{Numerical estimation}

The estimations of the instability growth rate,  \eqref{SRinstability:growthrate:largemass} for $\mu M\gg1$ and  \eqref{SRinstability:growthrate:smallmass} for $\mu M\ll1$, suggest that it becomes maximum at around $\mu M\sim1$. To prove this and find the maximum growth rate, we have to solve the ODEs for mode functions numerically. This numerical study was first done by Cardoso and Yoshida\cite{Cardoso.V&Yoshida2005} using the continued fraction method\cite{Leaver.E1985} and later extended to a larger parameter region by Dolan\cite{Dolan.S2007}. This method has been also used in calculating the quasi-normal mode frequencies of higher-dimensional black holes\cite{Cardoso.V&Yoshida2005,Konoplya.R&Zhidenko2011A} on the basis of master equations for perturbations\cite{Kodama.H&Ishibashi2003,Kodama.H&Ishibashi2004,Ishibashi.A&Kodama2003,Kodama.H2009,Ishibashi.A&Kodama2011} and establishing the SR instability of simply rotating adS black holes\cite{Kodama.H&Konoplya&Zhidenko2010,Kodama.H2008}. We briefly overview this numerical method in this subsection.

He first expanded the radial mode function $R$ as
\Eq{
R(r)=\frac{x^{-i\sigma}}{(r-r_-)^{\chi-1}} e^{-\sigma r} \sum_{n=0}^\infty a_n x^n,\qquad
x=\frac{r-r_+}{r-r_-},
}
where
\Eq{
q=\frac{2r_+(\omega-m\Omega_h)}{r_+-r_-},\quad
\sigma=(\mu^2-\omega^2)^{1/2},\quad
\chi = -(\mu^2-2\omega^2)/\sigma.
}
Inserting this into \eqref{ODE:radial}, we obtain the three term recurrence relation for the expansion coefficients $a_n$:
\Eq{
\alpha_n a_{n+1}+ \beta_n a_n + \gamma_n a_{n-1}=0.
}
with
\Eqrn{
\alpha_n=(n+1)(n+c_0),\quad
\beta_n =-2n^2 + (c_1+2)n + c_3,\quad
\gamma_n =n^2 + (c_2-3) n + c_4,
}
where $c_1,\cdots,c_4$ are constants dependent on $\omega,\sigma,m$ and $\Lambda_{lm}$.

The point is that under the assumption that $a_{n+1}/a_n$ tend to zero as $n\tend\infty$, we can solve this recurrence relation in terms of a continued fraction as
\Eq{
\frac{a_{n+1}}{a_n}=-\frac{\gamma_{n+1}}{\beta_{n+1}+\alpha_{n+1}\frac{a_{n+2}}{a_{n+1}} }
=-\frac{\gamma_{n+1}}{\beta_{n+1} - }\frac{\alpha_{n+1}\gamma_{n+2}}{\beta_{n+2} -}
 \frac{\alpha_{n+2}\gamma_{n+3}}{\beta_{n+3} - }\cdots.
}
Because $a_1/a_0=-\beta_0/\alpha_0$, this equation with $n=0$ gives the eigenvalue equation for $\omega=\omega_R + i\omega_I$:
\Eq{
\beta_0-\frac{\alpha_0\gamma_1}{\beta_1-}\frac{\alpha_1\gamma_2}{\beta_2-}\frac{\alpha_2\gamma_3}{\beta_3-}
 \cdots =0.
}

The convergence of this continued fraction is very rapid, and by truncating it at some appropriate level, we can easily obtain an approximate algebraic equation that enables us to determine $\omega^2$ with good accuracy.  As an example, we give the result of our numerical calculation for $l =m=1$ in Fig.\ref{fig:GrowthRate}.

\section{Axionic Instability of Astrophysical Black Holes}

Now, let us discuss what kind of astrophysical phenomena the superradiance instability provokes in realistic astrophysical systems with black holes. Description in this section are largely based on the axiverse paper\cite{Arvanitaki.A&&2010} and the paper by Arvanitaki and Dubovsky\cite{Arvanitaki.A&Dubovsky2011}, but it also contains new results obtained by us.

\subsection{Instability strip}

\begin{figure}
\centerline{
\includegraphics[height=6cm]{\FigDir/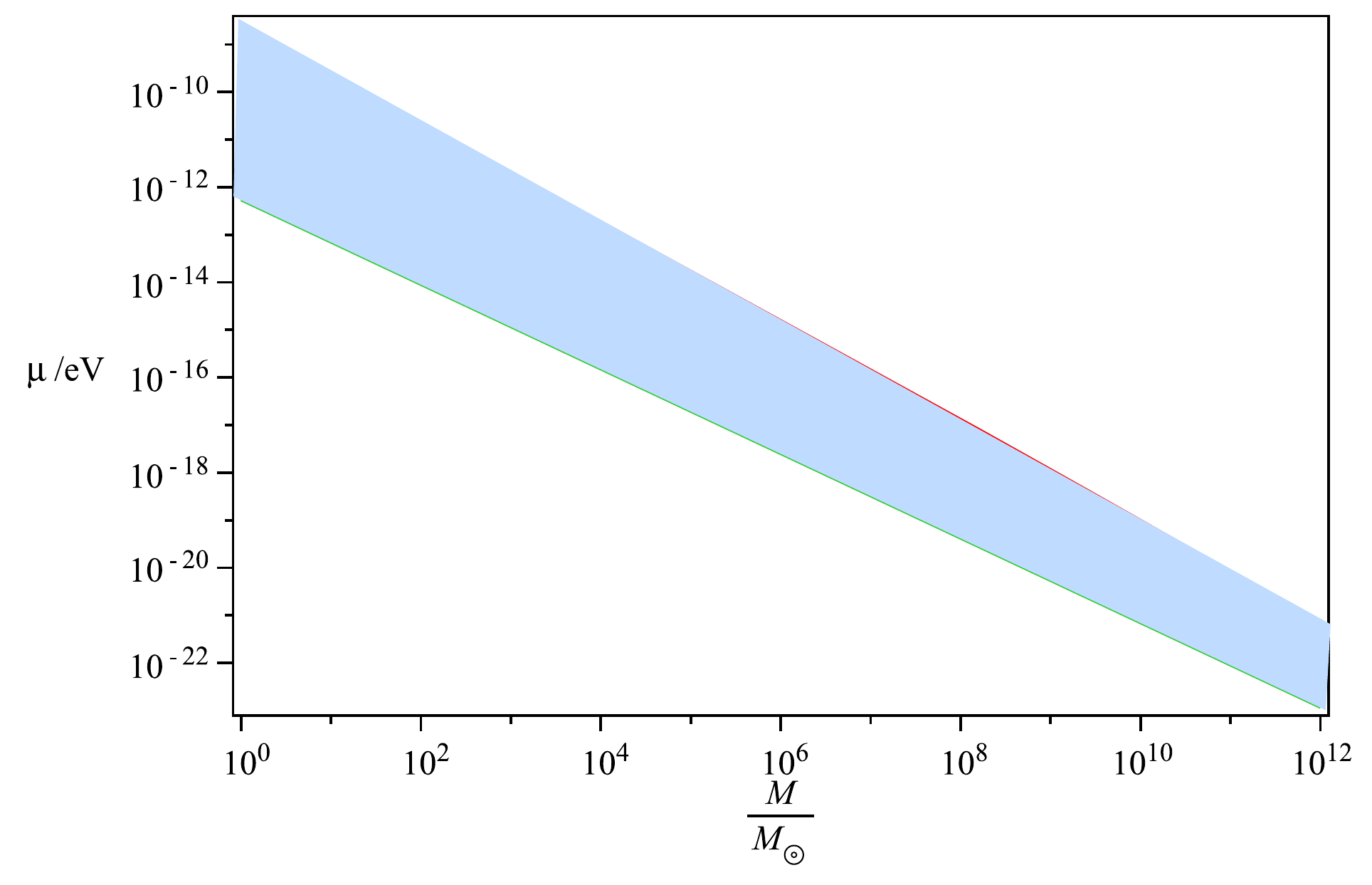}
}
\caption{The instability strip in the $\mu-M$ plane where the instability time scale is shorter than the cosmic age.}
\label{fig:InstabilityStrip}
\end{figure}

To start with, let us summarize the estimations obtained in the previous section. First, in asymptotic regions,  the growth time scale of the superradiance instability of a scalar field with mass $\mu$ around a Kerr black hole with mass $M$ and angular momentum $aM$ is approximately given by
\Eq{
\frac{\tau}{GM} \approx \cases{
                 10^7 e^{1.84 \alpha_g}  &; \alpha_g \gg1,\  a=1\\
                 24 \pfrac{a}{M}^{-1}\inpare{\alpha_g}^{-9} &; \alpha_g \ll1,
                 }
}
 where
\Eq{
\alpha_g := GM \mu =\frac{\mu}{1.34\cdot 10^{-10}{\rm eV}}\cdot \frac{M}{\Msun}
.
}
It becomes maximum at $\alpha_g\sim1$:
\Eq{
\tau_{\rm sr} \approx 0.2 \cdot10^7 GM;\quad
\alpha_g \simeq 0.44, \ a/M \simeq 0.999.
}

This result implies that if there exists a scalar field with mass $\mu\sim 10^{-10}{\rm eV}$ in nature, a cloud of the scalar field with a large amplitude will show up within one hour or so around any black hole with the solar mass due to the superradiant instability starting from tiny quantum fluctuations. This implies that the superradiant instability may provoke interesting astrophysical phenomena in reality, because the QCD axion $a$, whose existence is high probable taking account of the strong CP problem, can have a mass as small as $10^{-10}{\rm eV}$ if the PQ symmetry breaking scale is as large as $10^{16}{\rm GeV}$. This mass range is allowed if the initial amplitude during inflation is accidentally much smaller than a typical value.

Of course, this instability does not grow unboundedly because unstable modes have corotating angular momenta due to the superradiance condition and as a consequence, the central black hole loses its angular momentum by superradiance. Because the growth rate of the instability decreases rapidly with the decrease of the black hole angular momentum as shown in Fig.\ref{fig:GrowthRate}, and the instability stops when the black hole loses a non-negligible fraction of its angular momentum.

The most characteristic feature of this superradiance instability is the sensitivity of the growth rate on the masses of the scalar field and the black hole. Due to this, if we plot the region in the $\mu-M$ plane where the instability growth time scale  is shorter than the cosmic age ($\simeq 14$Gyr), we obtain a narrow strip as shown in Fig. \ref{fig:InstabilityStrip}. For example, if there exists a scalar field with mass $\mu\approx 10^{-14}{\rm eV}$, only systems with a black hole with mass in the range $10^2\Msun - 10^5 \Msun$ are affected and have smaller angular momenta than the other systems.

\subsection{G-atom}

\begin{figure}
\begin{minipage}{6cm}
\centerline{
\includegraphics[height=6cm]{\FigDir/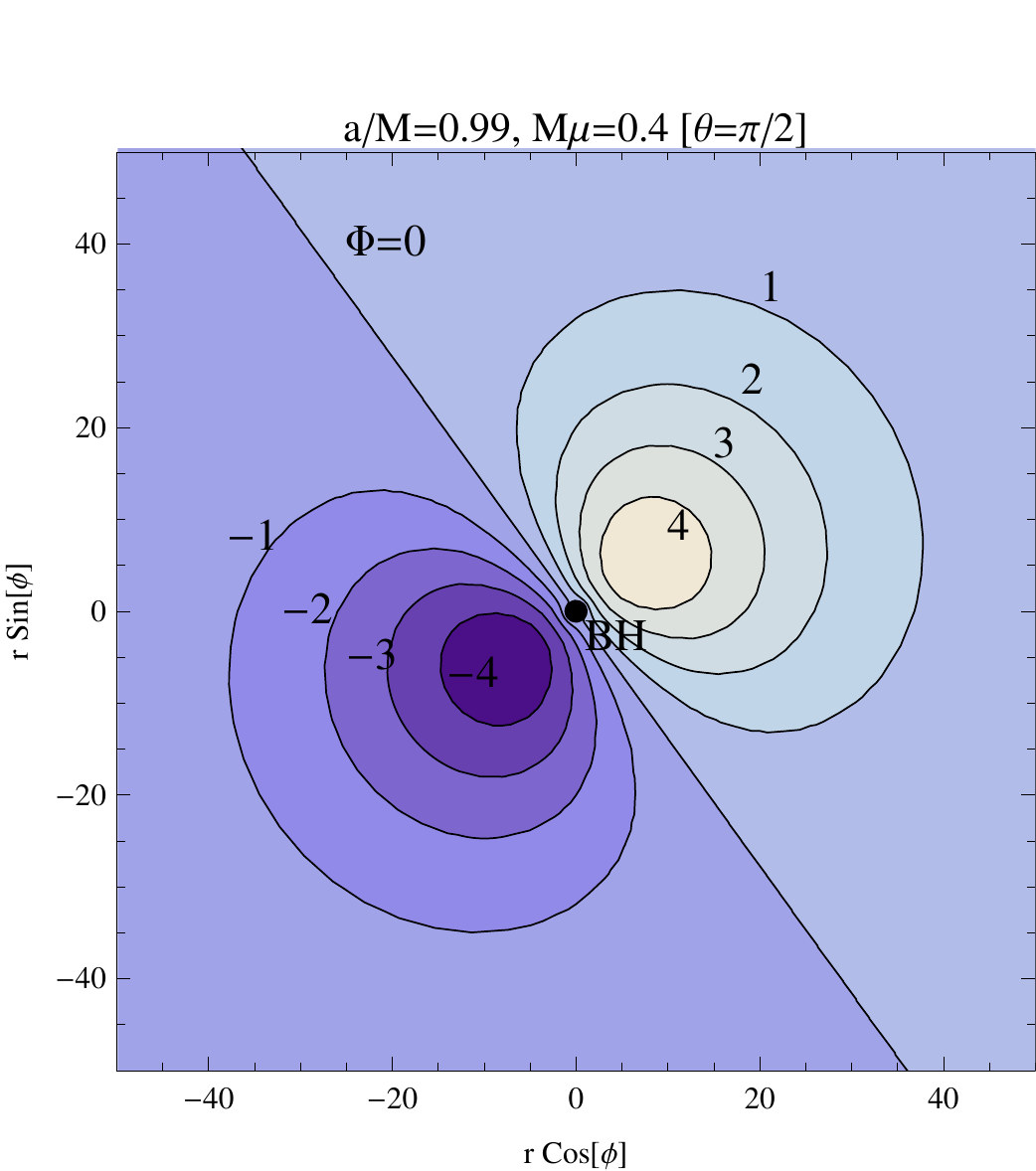}
}
\end{minipage}
\begin{minipage}{6cm}
\centerline{
\includegraphics[height=6cm]{\FigDir/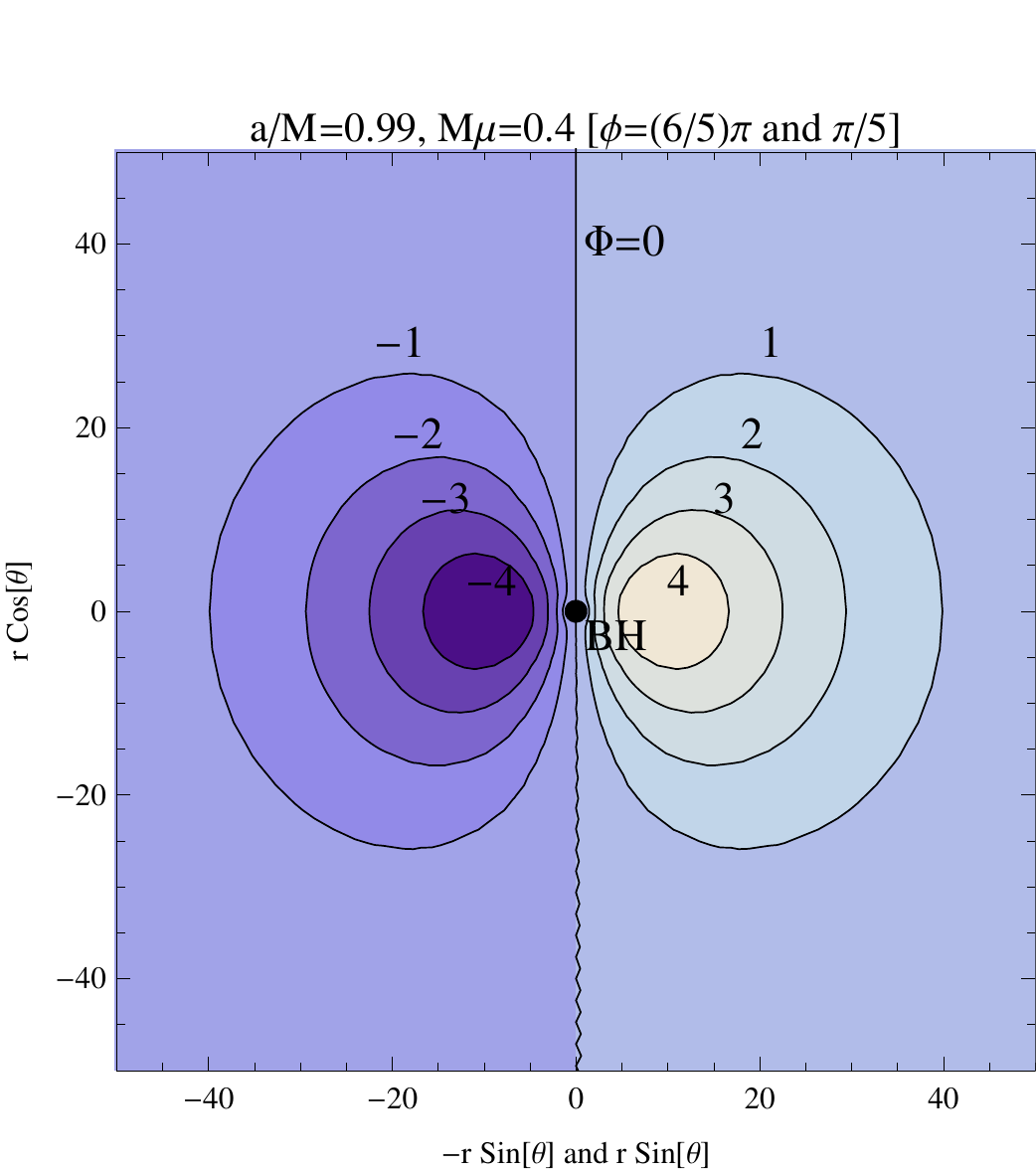}
}
\end{minipage}
\caption{The wave amplitude of the most unstable mode$(l=m=1)$ on the equatorial plane (the left panel) and on the vertical plane in the most extended direction (the right panel).}
\label{fig:GatomWF}
\end{figure}

Another peculiar feature of the superradiance instability of a black hole is the quantum nature of the phenomena. To see this, first recall that the instability occurs for bound states with $\omega\approx \mu$. If we approximate its energy levels by the hydrogen-atom type ones in the small mass case, from \eqref{hydrogenlevel}, we have
\Eq{
\omega_R^2 \simeq \mu^2 \inpare{1-\frac{\alpha_g^2}{2 n ^2}},\quad
\omega_R < m\Omega_h,
}
where $n=n'+l+1$ ($n'=0,1,2\cdots$). In general, for the most unstable mode with $n'\sim0$, $n$ can be estimated as
\Eq{
n \simeq l  \sim \frac{\mu}{\Omega_h}=\alpha_g \frac{2r_h}{a}.
}
Hence,   the mode is peaked outside the ergo region and  far from the horizon:
\Eq{
\frac{r}{R_g} \sim \frac{n^2}{\alpha_g^2}\sim 4\pfrac{ r_h}{a}^2
\then
\mu r \sim 4\alpha_g \pfrac{r_h}{a}^2 \sim 1
\label{quantumcondition}
}
This implies that the most unstable bound states are quantum for near extremal cases! Hence, the superradiance instability produces a G-atom (gravitational atom) consisting of the central black hole and a surrounding axion cloud in quantum states.  Fig. \ref{fig:GatomWF} illustrate the shape of this G-atom.

\subsection{Gravitational wave emission}

As shown in  Fig. \ref{fig:GatomWF}, the axion cloud of a G-atom is not spherically symmetric and rotating. Hence, it emits gravitational waves. The efficiency of this emission can be roughly estimated by the quadrupole formula. First, from \eqref{quantumcondition}, the cloud radius $r_c$ is around $r_c\sim M (l +1)^2/\alpha_g^2$, and we can assume that it is approximately Kepler rotating with the angular frequency $\Omega=(M/r_c^3)^{1/2}$. Then, from the quadrupole formula, the power of gravitational waves emitted from the axion cloud is
\Eq{
P =\frac{G}{45}|\dddot Q|^2  \sim \frac{G}{45}(r_c^2 \epsilon M)^2\Omega^6
 \sim \frac{\epsilon^2\alpha_g^{10}}{45 G(l +1)^{10}}=G \frac{N^2\alpha_g^{12}}{45(l +1)^{10} (GM)^4},
\label{G-atom:GW:quadrupole}
}
where $N$ is the number of axion quanta in the axion cloud, $\epsilon=\mu N/M$ is the ratio of the cloud mass to the black hole mass.

Here note that in general, the quadrupole formula estimates the gravitational wave emissions due to a slow change in the trajectories of source objects, which corresponds to transitions of the axion cloud from higher energy level to lower ones in the present problem. This implies that levels with $l \ge 2$ are relevant. Hence, the time scale for the axion cloud to lose a fraction $\epsilon$ of the gravitational energy of the black hole by quadrupole-type gravitational wave emission becomes
\Eq{
\tau_{\rm GW}\sim \frac{\epsilon M}{P} \approx \frac{45 GM (l +1)^{10}}{\epsilon \alpha_g^{10}}
\approx 10^{14} GM \pfrac{10^{-4}}{\epsilon}\pfrac{l +1}{3}^{10}\pfrac{0.44}{\alpha_g}^{10}.
}
From this, we obtain
\Eq{
{\tau_{\rm GW}} / {\tau_{\rm sr}} \approx 0.1 \times e^{-1.844(\alpha_g-2)}\inpare{2 /\alpha_g}^{10}.
}
This estimation implies that for $\alpha_g<2$, this time scale is much longer than the SR instability time scale, hence the quadrupole gravitational wave emissions do not stop the growth of the G-atom by instability. In contrast, for $\alpha_g>2$, $\tau_{\rm GW}$ becomes shorter than the SR instability time scale for $\epsilon\sim 10^{-4}$. This implies that for this parameter range, the SR instability stops growing when the cloud mass becomes $\epsilon M$ where the value of $\epsilon$ is determined by the condition $\tau_{\rm GW}=\tau_{\rm SR}$ because axions in unstable levels go to stable levels, provided that the other processes do not affect this balance.

From \eqref{G-atom:GW:quadrupole}, the observed amplitude of the GW metric perturbation is estimated as
\Eq{
h \approx 10^{-22} \pfrac{\epsilon}{10^{-4}} \pfrac{c^3}{GM\omega}\pfrac{100{\rm Mpc}}{d}\pfrac{M}{10^5\Msun}\pfrac{\alpha_g}{2}^5
 \pfrac{3}{l+1}^5
}
This indicates that massive black holes at galaxy centers produce gravitational waves with intensity observable by the next generation GW detection experiments such as the advanced LIGO, if there exists axions with mass in the range $10^{-15}{\rm eV}\lsim\mu\lsim 10^{-20}{\rm eV}$.

Of course, the estimation by the quadrupole formula may be too crude, and more precise estimations taking account of the quantum nature of the axionic cloud are required. Some preliminary analysis has been done in [\refcite{Arvanitaki.A&Dubovsky2011}], but a systematic study is yet to be done.

\begin{figure}
\centerline{
\includegraphics[width=12cm]{\FigDir/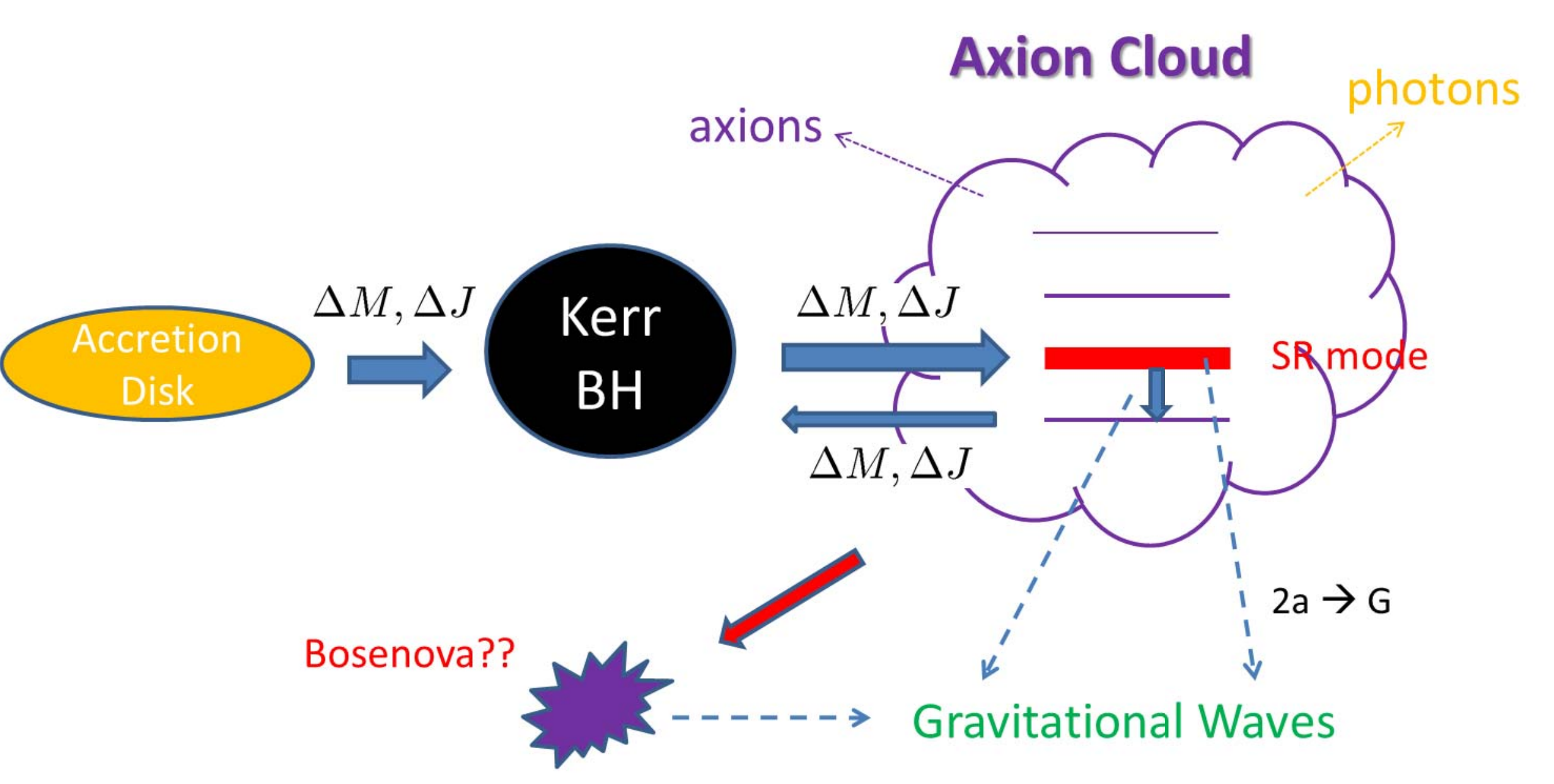}
}
\caption{The fate of an axion cloud around a black hole formed by instability.}
\label{fig:AxionCloud}
\end{figure}

\subsection{Bose nova}

In the discussions so far, we have neglected axion self-interactions. Because the axionic field grows coherently by superradiance instability, this approximation may become bad when the amplitude of the axionic field becomes large.

In order to estimate the effect of self-interactions, let us consider the axionic field $\phi$ with the action
\Eq{
S=\int d^4x\sqrt{-g} \insbra{-\frac{1}{2}(\nabla\phi)^2-\frac{\mu^2 f_a^2}{2}\sin^2\inpare{\phi/f_a}}.
\label{Axion:ExactAction}
}
Because the bound states of a G-atom is non-relativistic, we assume that the axionic field can be approximately written in terms of a slowly varing function $\psi$ as
\Eq{
\phi \simeq \frac{1}{\sqrt{2\mu}} \inpare{e^{-i\mu t} \psi + e^{i\mu t}\psi^*}.
}
Then, when $|\phi|/f_a\ll1$, we obtain a non-relativistic effective action
\Eq{
S_{\rm NR}= \int d^4x \insbra{ i\psi^* \pd_t \psi - \frac{1}{2\mu}\pd_i\psi \pd_i \psi^*
 -\mu \Phi_g\psi^*\psi +\frac{1}{16 f_a^2} (\psi^*\psi)^2 }.
}
The last term is the leading term of the interactions for small amplitudes. Because the corresponding energy is negative, it corresponds to an attractive force. Hence, it is expected that the axion cloud of the G-atom collapses due to this attractive force when the cloud becomes dense enough like the bosenova phenomena in the Bose-Einstein condensate as discussed by Arvanitaki and Dubovsky\cite{Arvanitaki.A&Dubovsky2011}.

\begin{figure}
\centerline{
\includegraphics[height=7cm]{\FigDir/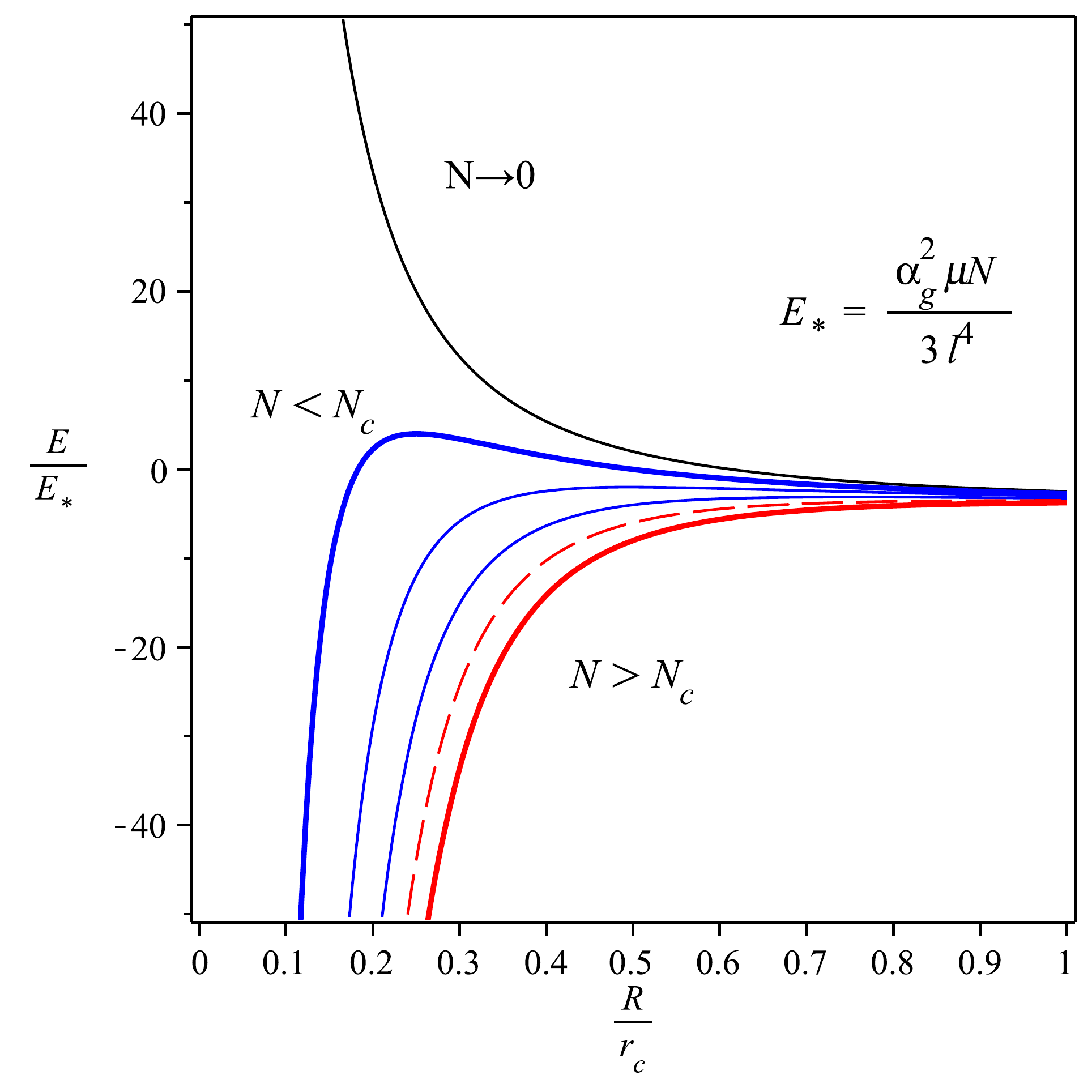}
}
\caption{The energy of an axion cloud as a function of the cloud size $R$}
\label{fig:AxionPotential}
\end{figure}

To see this, we approximate the total energy of the axion cloud as
\Eqrn{
E &=& \frac{V}{2\mu} \EXP{|\nabla\psi|^2} + \mu \Phi_g V\EXP{|\psi|^2}
 -\frac{V}{16f_a^2} \EXP{|\psi|^4}
\notag\\
 &\approx & \frac{N}{2\mu}\inpare{\frac{l ^2}{r^2}+\frac{1}{R^2}}
 -\frac{\alpha_g N}{r} - \frac{ N^2}{16f_a^2 R^3} ,
}
where $r$ is the distance of the cloud from the black hole center, and $R$ is the scale describing the extension of the cloud. By minimizing  $E$  w.r.t. $r$,  we get the Kepler radius of the axionic cloud:
\Eq{
r_c \approx \frac{l ^2}{\alpha_g \mu}
\then
E\approx \frac{N}{2\mu R^2} - \frac{N^2}{16 f_a^2 R^3}
 - \frac{\alpha_g N}{2r_c}.
}
This energy becomes maximum w.r.t. $R$  at  $R=R_m$. If  $R_m< r_c$,  the cloud with $R\sim r\sim r_c$ is stable,  while if $r_c< R_m$,  $E$ becomes a monotonically increasing function of $R$ in the range $R<r_c$ as shown in Fig.\ref{fig:AxionPotential}, and  the cloud initially with $R\sim r$ becomes unstable  and collapses rapidly. This happens when the total mass of the axion cloud exceeds the critical value given by
\Eq{
r_c < R_m \equivalent \mu N > \frac{16l ^2 f_a^2}{3\alpha_g \mu}
\equivalent
\epsilon =\frac{\mu N}{M}> \frac{l ^2 f_a^2}{\alpha_g^2 \mpl^2} \approx 10^{-4}.
}
%

\begin{figure}
\centerline{
\includegraphics[height=9cm]{\FigDir/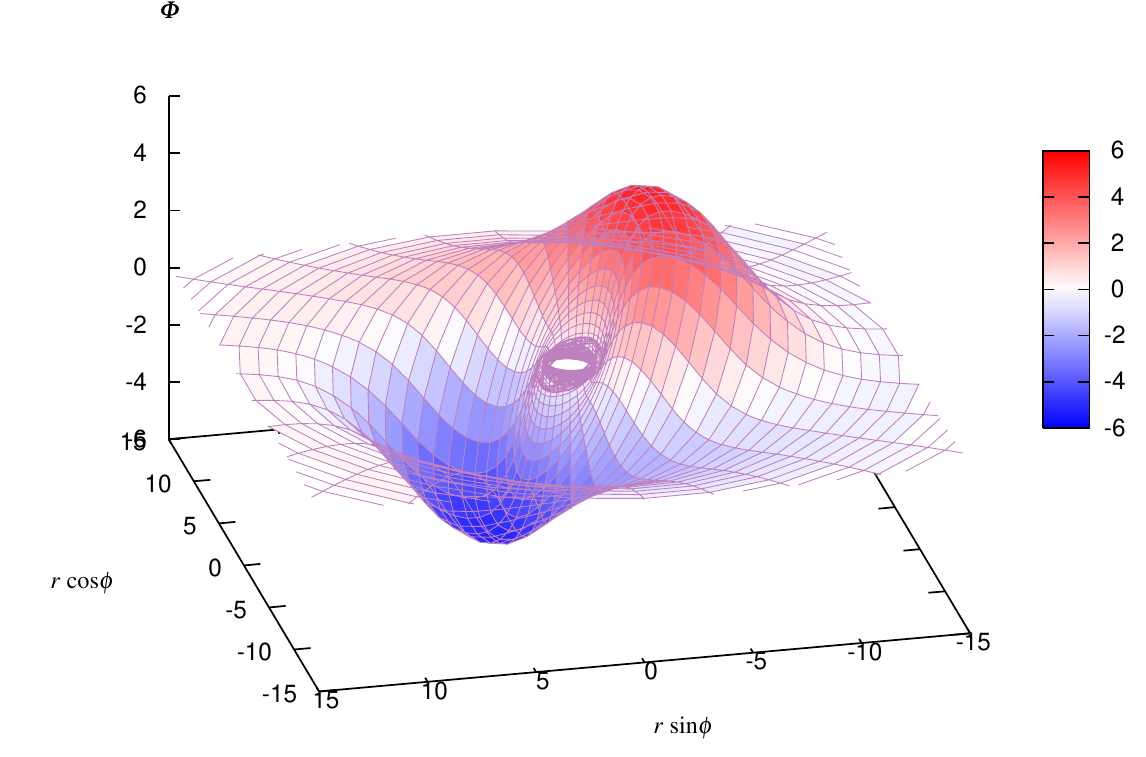}
}
\centerline{
\includegraphics[height=9cm]{\FigDir/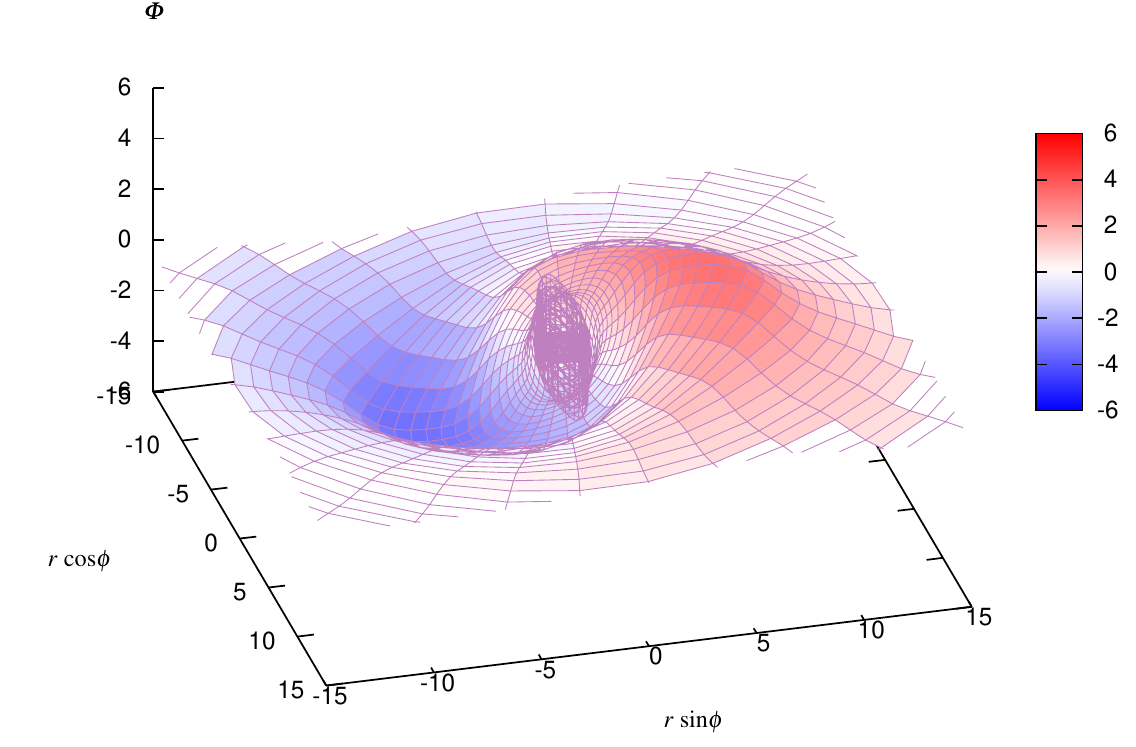}
}
\caption{Snapshots of a numerical simulation for bosenova collapse}
\label{fig:BoseNova}
\end{figure}

Here, it should be noted that the real self-interaction of the axion field is not quartic. When the amplitude increases, the higher-order terms with alternative signature come to contribute, and eventually when $|\phi|/f_a$ becomes of order unity, the full potential in proportion to $ \cos(\phi/\mu)$ should be used. Hence, it is a very interesting problem to find what really happens. To see this fate of a growing G-atom, we are now studying the evolution of an axion cloud around a Kerr black hole using the exact action \eqref{Axion:ExactAction}. According to the results we have obtained so far, the bosenova collapse of the G-atom really occurs when the amplitude of $\phi$ becomes comparable to $f_a$, i.e. when $\epsilon\sim 10^{-4}$. After the collapse, the large fraction of the energy deposited in the axion cloud falls toward the black hole as waves.  Figure \ref{fig:BoseNova} show two snapshots of the field amplitude on the equatorial plane before and after the bosenova collapse for a 3D numerical simulation of the axion evolution starting from the $2p$ mode for which the SR instability becomes maximum. The details of the results and physical analyses of them will be published elsewhere.

\section{Summary and Discussion}

In this article, we have overviewed the basic idea of the axiverse and its cosmophysical implications in order to show that superlight string axions can provide a new cosmophysical tool to probe the string theory/M-theory as the ultimate theory of nature. In particular, we have picked up the superradiance instability of astrophysical black holes as one of the most fascinating cosmophysical phenomena provoked superlight axions and given detailed accounts of its mechanism, the estimation of the instability growth rate and observational consequences.

Although many of the basic ideas of the axiverse and its cosmophysics are not new, the recognition of their importance is rather new. Furthermore, new ideas about axion cosmophysics are proposed every week. Therefore, there remain lots of problems to be studied systematically. For example,  in order to determine what really happens when the superradiance instability grows, in addition to gravitational wave emissions and the non-linear dynamics of axion clouds, we have to calculate various other processes such as the direct axion emissions due to non-linear interactions of axions and the radio emissions by the Primakov-type process taking account of the ubiquitous existence of strong magnetic fields in active black hole systems. In order to construct a model describing a realistic system, we also have to take into account the energy and angular momentum supply to the black hole by accretion.  Along with these cosmophysical investigations, it has also a crucial importance to construct string compactifications providing an effective four-dimensional cosmology consistent with all observations and experiments and calculate the axion spectrum on the basis of such compactifications. We can extract information on the ultimate theory only through the comparison of the string theory predictions obtained by such a first principle approach and the cosmophysical observations. Systematic investigations of both of these string theory and cosmophysics problems are now challenged under "Axiverse Project" in Japan.

\section*{Acknowledgements}

We would like to thank the Axiverse Project members, especially, Akihiro Ishibashi for valuable discussions. This work is supported by the MEXT Grant-in-Aid for Scientific Research on Innovative Areas (No. 21111006) and the JSPS Grant-in-Aid for Scientific Research (A) No. 22244030.


\end{document}